\documentclass[a4paper,fleqn,usenatbib]{mnras}

\usepackage{graphicx}	
\usepackage{amsmath}	
\usepackage{amssymb}	
\usepackage{multicol}        
\usepackage{bm}		
\usepackage{pdflscape}	
\usepackage{verbatim} 
\usepackage{siunitx}
\usepackage{xspace}

 \newcommand{\BY}{\mbox{\boldmath $Y$}}
 \newcommand{\bftheta}{\mbox{\boldmath $\theta$}}
 \newcommand{\bft}{\mbox{\boldmath $t$}}
\newcommand{\comm}[1]{#1}
\newcommand{\cas}[1]{#1}
\newcommand{\car}[1]{#1}
\newcommand{\gDor}{$\gamma$~Dor\xspace}
\newcommand{\percent}{~per~cent\xspace}
\newcommand{\Msun}{\,M$_\odot$\xspace}
\newcommand{\Table}[1]{Table~#1\xspace}
\newcommand{\gmodes}{g~modes\xspace}

\usepackage[T1]{fontenc}
\usepackage{ae,aecompl}

\usepackage{newtxtext,newtxmath}

\title[Asteroseismic modelling of \gDor stars]{Asteroseismic masses, ages, and
  core properties of $\gamma$~Doradus stars using gravito-inertial dipole modes 
  and spectroscopy }

\author[J.~S.~G. Mombarg et al.]{J.~S.~G. Mombarg$^{1}$\thanks{Contact: joey.mombarg@kuleuven.be}, 
T. Van Reeth$^{2,3,1}$, 
M.~G. Pedersen$^{1}$, 
G. Molenberghs$^{4,5}$, \newauthor
D.~M. Bowman$^{1}$, 
C. Johnston$^{1}$,
A. Tkachenko$^{1}$
and C. Aerts$^{1,6}$
\\
$^{1}$Institute of Astronomy, KU Leuven, Celestijnenlaan 200D, B-3001 Leuven, Belgium \\
$^{2}$Sydney Institute for Astronomy (SIfA), School of Physics, University of Sydney, New SouthWales 2006, Australia \\
$^{3}$\comm{Stellar Astrophysics Centre, Department of Physics and Astronomy, Aarhus University, Ny Munkegade 120, 8000 Aarhus C, Denmark \\}
$^{4}$I-BioStat, Universeit Hasselt, Martelarenlaan 42, B-3500 Hasselt, Belgium \\ 
$^{5}$I-BioStat, KU Leuven, Kapucijnenvoer 35, B-3000 Leuven, Belgium \\
$^{6}$Department of Astrophysics, IMAPP, Radboud University Nijmegen, PO Box 9010, 6500 GL Nijmegen, The Netherlands \\
}

\date{Accepted 2019 February 15. Received 2019 February 2019; in original form 2018 November 25}

\pubyear{2018}

\begin{document}
\label{firstpage}
\pagerange{\pageref{firstpage}--\pageref{lastpage}}
\maketitle

\begin{abstract}
\cas{
The asteroseismic modelling of period spacing patterns from gravito-inertial
modes in stars with a convective core is a high-dimensional problem.  We utilise
the measured period spacing pattern of prograde dipole gravity modes (acquiring $\Pi_0$), in
combination with the effective temperature ($T_{\rm eff}$) and surface gravity
($\log g$) derived from spectroscopy, to estimate the fundamental stellar
parameters and core properties of 37 $\gamma$~Doradus (\gDor) stars whose
rotation frequency has been derived from {\it Kepler\/} photometry. We make use
of two 6D grids of stellar models, one with step core overshooting and one with
exponential core overshooting, to evaluate correlations between the three
observables $\Pi_0$, $T_{\rm eff}$, and $\log g$ and the mass, age, core
overshooting, metallicity, initial hydrogen mass fraction and envelope mixing.
We provide multivariate linear model recipes relating the stellar parameters to
be estimated to the three observables ($\Pi_0$, $T_{\rm eff}$, $\log g$).  We
estimate the (core) mass, age, core overshooting and metallicity of \gDor stars
from an ensemble analysis and achieve relative uncertainties of $\sim\!10$\percent for
the parameters. The asteroseismic age determination allows us to conclude that efficient
angular momentum transport occurs already early on during the main sequence.  \car{We find that the nine stars with observed Rossby modes occur across almost the entire
main-sequence phase, except close to core-hydrogen exhaustion.} Future improvements of our work will come from the
inclusion of more types of detected modes per star, 
larger samples, and modelling of
individual mode frequencies.}
\end{abstract}





\section{Introduction}

The photometric data provided by space-based missions such as CoRoT
\citep{auvergne2009} and {\it Kepler} \citep{koch2010} have heralded a new era
for asteroseismology. Here, we are concerned with gravito-inertial
asteroseismology, i.e., the study of gravity modes (\gmodes) in rotating intermediate-mass
stars.  Such modes are subject to both the Coriolis force and buoyancy as
restoring forces.  While CoRoT data led to the first discoveries of period
spacings of gravity-mode pulsators in the core-hydrogen burning phase
\citep{degroote2010,papics2012}, it did not allow \comm{the identification of} the \comm{angular} degree of
these detected oscillations without ambiguity. Secure mode identification had to await \comm{nearly uninterrupted time series photometry with at least a factor ten longer time base}, such as that assembled with the {\it Kepler\/} space telescope \citep{borucki2010}. Meanwhile, lots of progress has been made on the
observational side in this topic over the past few years with firm detections of period spacing patterns reported in \cite{papics2014,papics2015,kurtz2014,saio2015,VanReeth2015b, schmid2015,VanReeth2016, murphy2016, guo2016,ouazzani2017,saio2017,saio2018,szewczuk2018} and \citet{li2018}.

As shown in these papers, we have now reached the stage where tens of gravity-mode
frequencies have been measured in these pulsators with sufficient precision to identify their mode
degree from the 4-year nominal {\it Kepler\/} light curves and hence to start
testing and improving stellar structure theory of intermediate-mass stars.

Forward modelling applications to a few stars have shown the need of core
overshooting and envelope mixing in both B-type and F-type gravity-mode
pulsators, in order to be able to explain the measured mode trapping properties
\citep{moravveji2015,moravveji2016, schmid2016}.  Moreover, the
near-core rotation rates derived from the \gmodes \citep{VanReeth2016,ouazzani2017,VanReeth2018} have revealed shortcomings
in stellar evolution theory in terms of angular momentum transport, both
during the core-hydrogen burning phase
\citep{rogers2015,aerts2017,townsend2018,ouazzani2018} and in the red giant
phase \citep{mosser2015,eggenberger2017,gehan2018} -- see \citet{aerts2019} for a recent extensive review.

Since the evolution of a star is greatly affected by the stellar rotation
profile \citep{maeder2009}, the opportunity to calibrate theoretical stellar
models from empirically derived angular momentum distributions in stellar
interiors at different evolutionary stages is of major importance.  In order to
compute these angular momentum distributions, the mass and radius of the star as
a whole and of its convective core have to be estimated, as well as the star's
age. Given that forward asteroseismic modelling of intermediate-mass stars is a
high dimensional problem \citep[+6D,][]{aerts2018}, a \comm{robust statistical} methodology is
needed.

In this paper, we explore the feasibility of estimating the most important
stellar parameters, i.e., (core) mass, core overshooting, initial hydrogen mass fraction,
metallicity, \comm{the amount} of envelope mixing, and age (or age-proxy) of $\gamma$~Doradus
(henceforth \gDor) stars. While these pulsators of spectral type \comm{late-A to early-F}
were already known to have \gmodes excited by a flux blocking mechanism
from ground-based studies
\citep[e.g.,][]{kaye1999,guzik2000,dupret2005,cuypers2009,bouabid2013}, the derivation of
g-mode period spacings had to await the long-term uninterrupted {\it Kepler\/}
data \citep[e.g.,][]{bedding2015, VanReeth2015b,VanReeth2015a}.

\cite{VanReeth2016, VanReeth2018} and \cite{ouazzani2017} have developed methods to infer the
near-core rotation frequency from the slope of the measured period spacing
pattern of \gDor stars. These authors applied their methods to ensembles
of 40 \comm{and 4 stars}, respectively, of such gravito-inertial pulsators. Here, we consider
the sample by \citet{VanReeth2016} because 37 of these stars have been monitored
with high-resolution spectroscopy \citep{tkachenko2013} and the asymptotic
period spacing from their prograde sectoral dipole modes, effective \comm{temperatures}, surface \comm{gravities}
and metallicity have been determined in a homogeneous way.

We explore the modelling capacity of the combined \cas{seismic parameter} for \comm{dipole modes (in the non-rotating limit)},
$\Pi_0$, and the spectroscopically derived effective temperature, $T_{\rm eff}$ and surface gravity, $\log g$, as a major simplification \comm{of typical} forward modelling that is based on the fitting
of all the individual g-mode frequencies.  In \comm{our approach}, we first derive
correlations between the \cas{the seismic parameter $\Pi_0$}, the effective temperature and surface gravity
on one hand, and \comm{the correlation between these three observables and the stellar parameters varied in two 6D stellar model grids on the other hand}.  Previous studies in the literature have derived correlations between the
stellar mass, metallicity, and step overshooting for low-order modes in
$\beta$~Cep stars \citep[e.g.,][]{briquet2007, walczak2013} and for high-order
g-modes in a {\it Kepler\/} slowly pulsating B-type star \citep{moravveji2015}. \comm{In Section \ref{sec:stat_mod}},
we \comm{investigate} correlations between mass, \comm{central} and initial hydrogen fraction,
metallicity, \comm{the amount} of envelope mixing, and the
mass and \comm{radius} of the convective core, by means of linear multivariate
regression \comm{to investigate the correlations between the parameters in a simple manner}. We do this for both a step and an exponential core overshooting
formalism since it was recently shown that \gmodes potentially allow \comm{for} these two overshooting prescriptions \comm{to be distinguished} \citep[][]{pedersen2018}. In
particular, we also investigate how well a benchmark model -- based on one of these
two core overshooting prescriptions -- can be approximated by a model based on the
other overshooting prescription for the mass range of \gDor stars \comm{in Section \ref{sec:BM_overshoot}}.

Armed with the knowledge of the correlation structure in the two 6D model grids,
we explore the capacity of \comm{these diagnostics} $(\Pi_0, T_{\rm eff}, \log g)$ for
parameter estimation from seismic modelling, using the methodology based on
maximum likelihood estimation developed in \citet{aerts2018}. \comm{The results of our forward modelling are presented in Section \ref{sec:Mahalanobis}. In Section \ref{sec:err_est}, we describe our methodology of determining uncertainties from ensemble modelling and we conclude in Section \ref{sec:conclusions}.}

\section{Properties of 6D asteroseismic model grids}
\label{sec:stat_mod}

\subsection{Gravity-mode period spacings}
\cas{In the framework of the traditional approximation of rotation (TAR; \citealt{eckart1960, unno1989}), 
the period of a high-order g-mode can be well approximated by
\begin{equation}
    P_{\rm co} \approx \frac{\Pi_0}{\sqrt{\lambda_{l,m,s}}}(n_{\rm g} + \alpha_{\rm g}),
\end{equation}
with $\lambda$ the eigenvalue of the Laplace tidal equation, depending on the
mode geometry (spherical degree $l$ and azimuthal order $m$) and the spin
parameter $s = 2f_{\rm co}/f_{\rm rot}$ 
\citep[cf.,][]{Townsend2003, bouabid2013,VanReeth2018}. The phase
term $\alpha_{\rm g}$ depends on the stellar structure
and
\begin{equation}
    \Pi_0 \equiv 2\pi^2\left(\int_{\rm gc} \frac{N}{r} {\rm d}r\right)^{-1}.
    \label{eq:Pi_0}
\end{equation}
Here, the quantity $N$ is the Brunt-V\"ais\"al\"a frequency and $N/r$ is 
integrated over the
gravity-mode cavity indicated as `gc'.
An example of such a cavity is shown in Fig.\,\ref{fig:mode_cavity} in the Appendix.
The asymptotic period spacing of \gmodes, i.e., the difference in period between two
modes of 
consecutive radial order and same angular degree, in the case of a chemically homogeneous, non-magnetic
star \comm{is} defined in the corotating frame as
\begin{equation}
    \Delta P_{\rm co} \approx \frac{\Pi_0}{\sqrt{\lambda_{l,m,s}}}.
    \label{Eq:dP_co}
\end{equation}
Depending on the nature of the mode and on the value of the spin parameter, 
the value of $\lambda_{l,m,s}$ can be approximated by simple analytical expression
\citep[cf.,][Fig.\,A1]{Townsend2003, saio2017}. The 
spin parameter for the gravito-inertial modes and Rossby modes of the stars
in our sample ranges from 1 to 30 \citep[cf., Fig. 2 of][]{aerts2017}. Given this broad range, it is not obvious to resort to analytical approximations for $\lambda_{l,m,s}$ in the case of $\gamma\,$Dor stars.
This motivated \citet{VanReeth2016} to work with numerical solutions of the Laplace tidal equation.

Comparison between the measured and theoretically predicted gravity-mode periods
requires the  transformation of the periods to an inertial frame of reference. In the
case of a uniform stellar rotation with frequency $f_{\rm rot}$, 
the periods of the oscillation modes in an inertial frame can be computed as
\begin{equation}
    P_{\rm inert} = \frac{1}{f_{\rm co} + mf_{\rm rot}},
\end{equation}
where $f_{\rm co}$ is the mode frequency in the corotating frame. We adopt the
convention that prograde modes correspond to $m$~\textgreater~0. 

In the forward modelling applied here, 
we rely on the observational estimates of $\Pi_0$ and of $f_{\rm
  rot}$ as determined by \citet{VanReeth2016,VanReeth2018}. We therefore  compute the quantity $\Pi_0$
from theoretical stellar evolution models. 
The diagnostic power of the Brunt-V\"ais\"al\"a frequency and the $\Pi_0$ of} \comm{\gmodes are} well known and \comm{have} already been exploited 
asteroseismically
since ground-based multi-site white dwarf asteroseismology \citep{Winget1991, brassard1992}. The \gmodes of these compact objects have periods of order a
few to ten minutes, such that beating patterns can be covered in observing runs
lasting \cas{several} weeks. Moreover, the \comm{rotation period} of such stars is of order days,
such that the rotational effect on the pulsation modes can be treated from a
perturbative approach.  The models and \gmodes in white dwarfs are quite
different from those of young stars, hence the parameters that can be deduced
from them, as well as the mode trapping properties differ accordingly.  From an
observational point of view, the application and exploitation of g-mode
asteroseismology to core-hydrogen burning stars had to await long-term
uninterrupted {\it Kepler\/} photometry, given that the modes have periodicities
of order days and beating patterns of years. 
Moreover, the \gmodes and rotation \comm{rates in these stars}
have similar periodicities, such that rotation cannot be treated perturbatively,
except for a few ultra-slow rotators. This requires dedicated
asteroseismic modelling tools suitable to interpret the measured
frequencies. Here, we explore and apply aspects of the methodology developed specifically for gravito-inertial modes by \citet{aerts2018}. We first highlight relevant properties of the asteroseismic
grids upon which we rely and subsequently exploit the probing power of the \comm{three observables}
\cas{($\Pi_0$, $T_{\rm eff}$, $\log g$).}

\subsection{Correlations among the stellar model parameters}
\cite{VanReeth2016} computed two extensive grids of non-rotating stellar modes with the
MESA stellar evolution code \citep[r7385,][and references therein]{paxton2018}. These models vary in stellar mass ($M_{\star}$),
metallicity ($Z$), diffusive envelope mixing ($D_{\rm mix}$, \comm{constant throughout the radiative zone}), the extension of
the core overshoot region $\alpha_{\rm ov}/f_{\rm ov}$ (i.e., for
step/exponential overshoot, expressing the mean free path of a convective fluid element in a radiative region in terms of the local pressure scale
height), the initial ($X_{\rm ini}$) and \comm{normalised} central hydrogen content 
($X_{\rm c}' = X_{\rm c}/X_{\rm ini}$). The latter is a proxy for
the evolutionary stage, hence for the age of the model.   In this paper, these two
grids of evolution models were extended to lower mass compared to \citet[][]{VanReeth2016} \comm{-- a lower limit of 1.2\Msun instead of 1.4\Msun}.  For the
models with a convective envelope, we did not consider any undershooting, because this is not important to assess the probing power of high-order \gmodes.

In \Table{\ref{tab:grid}}, an overview is given of the parameters varied across
the two grids and respective step sizes.  \cas{The efficiency of convection, in
  our model grids parameterized by the mixing length parameter
  $\alpha_{\rm MLT}$, was kept fixed at 1.8 throughout the grids. As this was
  done without further discussion in \citet{VanReeth2016}, we point out here
  that this parameter does influence the three quantities $\Pi_0$,
  $T_{\rm eff}$, and $\log g$. This is particularly the case for the lower-mass
  stars in the model grids, since these have the larger convective envelope and
  are hence affected more by the treatment of convection than the higher-mass
  stars.  Using table~2 from \citet{viani2018} with the appropriate $\alpha_{\rm MLT}$ value, we made an estimate for the
  range of the mixing length parameter for the stars in our sample, resulting in
  $\alpha_{\rm MLT}\in [1.5,1.9]$.  Varying $\alpha_{\rm MLT}$ between 1.5 and
  1.9, we find that only for the most evolved low-mass stars do the differences in
   $T_{\rm eff}$ become comparable to typical
  observational uncertainties, while for $\Pi_0$ and $\log g$ the differences are always negligible. We illustrate this in Figs\,\ref{fig:Pi0}--\ref{fig:logg} in Appendix\,\ref{appendix:alpha_MLT} for a low-mass and
  high-mass \gDor model.  We do stress that $\alpha_{\rm MLT}$ has a major
  influence in forward modelling based on fitting individual g-mode
  frequencies \citep[][table\,2]{aerts2018}, but that this dependency plays an
  inferior role compared to the other parameters varied in our grids for forward
  modelling based on $\Pi_0$.  }

\cas{ For the modelling done here, we also transformed to the parameter
  $X_{\rm c}' = X_{\rm c}/X_{\rm ini}\in [0,1]$, i.e., to the fraction of the
  main-sequence duration, as a proxy for the age, rather than using $X_{\rm c}$
  itself as in \citet{VanReeth2016}.  In this way, the span of $X_{\rm c}'$ in the grids is
  the same for all model tracks ranging} from the zero-age main sequence (ZAMS;
$X_{\rm c}' = 1$) to the terminal-age main sequence (TAMS; $X_{\rm c}' = 0$). \comm{After
  the onset of core-hydrogen burning at the ZAMS, a model requires a few
  iterations to return to hydrostatic equilibrium} and therefore we do not
\comm{consider any} models with $X_{\rm c}' > 0.99$. In total, the step overshoot grid
contains \num{1530} evolutionary tracks with \num{819774} stellar models and the
exponential overshoot grid contains \num{2295} evolutionary tracks with
\num{1300590} stellar models.

\cas{For each stellar model in the grids, $\Pi_0$ was computed, along with the
  effective temperature $T_{\rm eff}$ and surface gravity $\log g$, as well as
  the mass $M_{\rm cc}$ and radius $R_{\rm cc}$ of the convective core.}
\cas{We refer to Table\,\ref{tab:MLE_Grid} in Appendix\,\ref{AppendixA1}, where
  we assembled the measured values of $\Pi_0$ from {\it Kepler\/} data as
  determined by \citet{VanReeth2018} under the assumption of rigid rotation. For
  the stars with both gravito-inertial prograde dipole modes and Rossby modes,
  these are improved values compared to those in \citet{VanReeth2016}.  The
  observational estimate of $\Pi_0$ for all these stars was deduced along with
  estimation of the near-core rotation frequency $f_{\rm rot}$ from the slope of
  the measured period spacing patterns.  The relative observational errors for
  $\Pi_0$ range from 0.2 to 26\percent for the various sample
  stars. Below, we compare these observables with the theoretical predictions
  for $\Pi_0$ computed for each of the grid models.} We recall explicitly that
  \citet{VanReeth2016} carefully omitted modes that are trapped to estimate
  $\Pi_0$ and $f_{\rm rot}$ from the data (cf., their figs\,4 and 9).
  \cas{These two observed quantities, along with the mode identification, were
    derived using the TAR to compute $\lambda_{l,m,s}$ with the pulsation code
    GYRE \citep{townsend2013,townsend2018} for each of the models in the two
    extensive 6D grids, which cover the relevant parameter ranges of $\gamma\,$Dor
    stars. In this way, our observational estimation of $\Pi_0$ and $f_{\rm
      rot}$ does not depend on particular choices of these model parameters.}

\begin{table}
    \centering
    \begin{tabular}{c|ccc}
        \hline        
        Parameter & Lower boundary & Upper boundary & Step size \\
        \hline
        \multicolumn{4}{c}{\textbf{Step overshoot}} \\
        \hline
        $M_{\star}$ & 1.20\Msun & 2.00\Msun & 0.05\Msun  \\
        $Z$ & 0.010 & 0.018 & 0.004 \\
        $\log D_{\rm mix}[$cm$^2$s$^{-1}]$ & -1 & 0 & 1 \\
        $\alpha_{\rm ov}$  & 0.01 & 0.300 & 0.075 \\
        $X_{\rm ini}$ & 0.69 & 0.73 & 0.02 \\
        $X_{\rm c}'$ & 0 & 0.99 & \textless0.007 \\
        \hline
        \multicolumn{4}{c}{\textbf{Exponential overshoot}}\\
        \hline
        $M_{\star}$ & 1.20\Msun & 2.00\Msun & 0.05\Msun  \\
        $Z$ & 0.010 & 0.018 & 0.004 \\
        $\log D_{\rm mix}[$cm$^2$s$^{-1}]$ & -1 & 1 & 1 \\
        $f_{\rm ov}$  & 0.001 & 0.0300 & 0.0075 \\
        $X_{\rm ini}$ & 0.69 & 0.73 & 0.02 \\
        $X_{\rm c}'$ & 0 & 0.99 & \textless 0.007 \\
        \hline
    \end{tabular}
    \caption{Range of the model grids for which $\Pi_0$, $T_{\rm eff}$, and 
$\log g$ have been computed. }
    \label{tab:grid}
\end{table}

\begin{figure*}
    \centering
    \includegraphics[width = \textwidth]{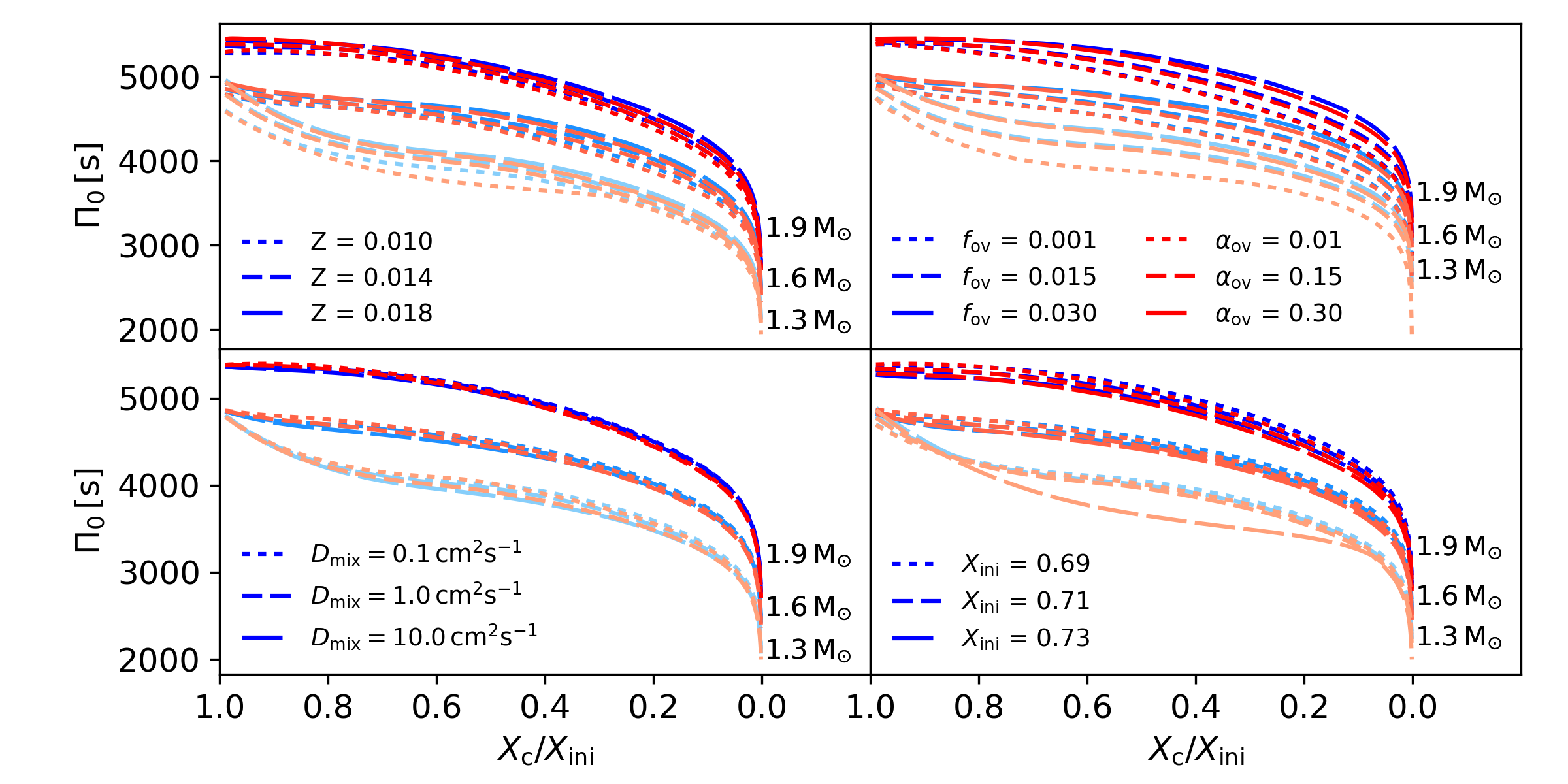}
    \caption{\cas{Evolution of $\Pi_0$ along the
      main sequence for changing
      metallicity (top left), amount of core overshooting (top right), \comm{the amount} of mixing in the radiative zone (bottom left) and initial hydrogen mass fraction (bottom right), for a 1.3-, 1.6- and
      1.9-${\rm M}_\odot$ star (lightest colour to darkest)}. The \comm{blue} lines correspond to an exponential
      overshooting prescription and the red lines to a step overshooting prescription, both of
      which consider the radiative temperature gradient in the overshoot zone. When a
      parameter is not being varied, it is set at $Z = 0.014$, $\alpha_{\rm ov} = 0.225$, $f_{\rm ov} = 0.0225$, $D_{\rm mix} = \SI{1.0}{\cm\squared\per\second}$ and $X_{\rm ini} = 0.71$.}
    \label{fig:dP-Xc}
\end{figure*}

We recall that mixing in the radiative zone had to be introduced in addition to
core overshooting in models of B-type pulsators to fit the frequencies of
trapped \gmodes \citep{moravveji2015,moravveji2016}.  In Fig.\,\ref{fig:dP-Xc},
the evolution of $\Pi_0$ as a function of the stellar age is illustrated, where
either $X_{\rm ini}$, $Z$, $\alpha_{\rm ov}/f_{\rm ov}$ or $D_{\rm mix}$ is
being varied, while the three remaining parameters are kept fixed, for masses
1.3, 1.6, and 1.9\Msun. This gives a good visual representation of the
dependencies of $\Pi_0$ on these four model parameters in the grids. It can be
seen that there is no unique \comm{monotonic} relation between mass, age and
$\Pi_0$ because the core overshooting, metallicity and initial hydrogen (in this
\comm{descending} order of importance) do have an effect larger than the typical
measurement uncertainty of $\Pi_0$ deduced from the {\it Kepler\/} light curves
\citep[cf.,][for a thorough discussion of theoretical uncertainties on g-mode
frequencies]{aerts2018}.  The envelope mixing does not influence the period
spacing values as much as the other three parameters. This is entirely as
expected, given that the \gmodes mainly probe the core \comm{overshooting},
while they are hardly affected by the envelope mixing. However, the mixing at
the interface between the overshoot zone and the bottom of the radiative
envelope is responsible for the details of the mode trapping and has to be
considered when performing frequency fitting \comm{as opposed to} matching of
the asymptotic period spacing
\citep[cf.,][]{moravveji2016,VanReeth2016,pedersen2018}.

\subsection{Linear statistical model for \texorpdfstring{$\Pi_0$}{Pi0}}

\cas{ Following problem set 3 in \citet{aerts2018},
we construct a statistical model for $\Pi_0$ based on the
stellar input parameters of the model grids}
by performing a linear multivariate regression,
adopting the following form 
\begin{equation}
\Pi_0 = \beta_0 + \beta_1 \alpha_{\rm ov} + \beta_2 D_{\rm mix} + \beta_3 M_{\star} + \beta_4 Z + \beta_5 X_{\rm ini} + \beta_6  X_{\rm c}' \cas{+ \epsilon}, 
    \label{fit6}
\end{equation}
\cas{where $\epsilon$ is the residual and the}
 coefficients $\beta_i$ are computed according to an 
ordinary least-squares regression,
\begin{equation}
    \bm{\beta} = (\bm{X}^\top \bm{X})^{-1}\bm{X}^\top \bm{Y}.
\end{equation}
\cas{Here,} $\bm{Y}$ is a vector with length equal to the number of grid points $i = 1,\ldots, q$,
containing all values of $\Pi_{0,i}$ and 
\begin{equation}
\bm{X} = 
 \begin{pmatrix}
  1 & \alpha_{{\rm ov},1} & D_{{\rm mix},1} & M_{\star,1} & Z_1 & X_{{\rm ini},1} & X_{\rm c,1}'  \\
  1 & \alpha_{{\rm ov},2} & D_{{\rm mix},2} & M_{\star,2} & Z_2 & X_{{\rm ini},2} & X_{\rm c,2}'  \\
  \vdots & \vdots & \vdots & \vdots & \vdots & \vdots & \vdots  \\
  1 & \alpha_{{\rm ov},q} & D_{{\rm mix},q} & M_{\star,q} & Z_q & X_{{\rm ini},q} & X_{{\rm c},q}'  \\
 \end{pmatrix}.
\end{equation}
For the step overshoot grid, $q = \num{819774}$, while for the exponential overshoot grid, $q = \num{1300590}$. 

\cas{The residual term $\epsilon$ captures the non-linearity in $\Pi_0$ as it
  occurs in the model grids. We point out that both the coefficients $\beta_i$
  and $\epsilon$ depend on the choices of microphysics that went into the
  computation of the stellar models. This concerns opacity tables, the equation
  of state, nuclear reactions and chemical mixtures. The effect of these
  choices on g-mode pulsation frequencies cannot be represented by a few
  simple model parameters, as discussed in detail in sections 2 to 4 in 
  \citet{aerts2018}. That paper's table\,2 also includes a detailed quantification and
  hierarchical ordering of the effect of the choices of micro- and macro-physics
  when performing forward asteroseismic modelling of main-sequence stars.}

\begin{table}
    \centering
    \begin{tabular}{lcc}
        \hline
        \, & Step overshoot & Exponential overshoot\\
        \hline
        $\beta_0\,({\rm Offset})$ &  1095.74$\,\pm\,$12.64 & 1112.30$\,\pm\,$10.05 \\
        $\beta_1\,(\alpha_{\rm ov})$ &  -258.2$\,\pm\,$3.6 & -3076.99$\,\pm\,$30.43\\
        $\beta_2\,(D_{\rm mix})$ & -53.0$\,\pm\,$0.6 & -7.70$\,\pm\,$0.05 \\
        $\beta_3\,(M_{\star})$ & 520.2$\,\pm\,$2.0 & 631.0$\,\pm\,$1.5 \\
        $\beta_4\,(Z)$ & 32236.60$\,\pm\,$87.77 & 32201.55$\,\pm\,$70.09 \\
        $\beta_5\,(X_{\rm ini}) $& 456.17$\,\pm\,$17.74 & 236.57$\,\pm\,$14.07\\
        $\beta_6\,(X_{\rm c}')$ & 2620.4$\,\pm\,$3.4 & 2716.8$\,\pm\,$2.7\\
        $\beta_7\,(M_{\rm cc})$ & 22176.29$\,\pm\,$33.18 & 21397.18$\,\pm\,$25.16 \\
        $\beta_8\,(R_{\rm cc})$ & -24236.95$\,\pm\,$65.79 & -25246.43$\,\pm\,$50.06 \\
        \hline
    \end{tabular}
    \caption{\cas{Regression coefficients from an ordinary
      least squares fit for the best statistical model in Eq.\,(\ref{fit7}) according to the BIC for $\Pi_0$ (s).}}
    \label{tab:betas_dP}
\end{table}

\begin{table}
    \centering
    \begin{tabular}{lcc}
        \hline
        \, & Step overshoot & Exponential overshoot\\
        \hline
        $\beta_0\,({\rm Offset})$ &  7995.63$\,\pm\,$11.70 & 7872.1$\,\pm\,$9.1 \\
        $\beta_1\,(\alpha_{\rm ov})$ &  -393.9$\,\pm\,$3.3 & -6509.54$\,\pm\,$27.48\\
        $\beta_2\,(D_{\rm mix})$ & - & -0.48$\,\pm\,$0.04 \\
        $\beta_3\,(M_{\star})$ & 2508.7$\,\pm\,$1.9 & 2351.6$\,\pm\,$1.4 \\
        $\beta_4\,(Z)$ & -87369.35$\,\pm\,$81.21 & -84919.61$\,\pm\,$63.29 \\
        $\beta_5\,(X_{\rm ini}) $& -6879.91$\,\pm\,$16.41 & -6606.00$\,\pm\,$12.71\\
        $\beta_6\,(X_{\rm c}')$ & -792.5$\,\pm\,$3.1 & -720.1$\,\pm\,$2.5\\
        $\beta_7\,(M_{\rm cc})$ & -11630.93$\,\pm\,$30.69 & -7662.64$\,\pm\,$22.72 \\
        $\beta_8\,(R_{\rm cc})$ & 39468.86$\,\pm\,$60.87 & 36305.27$\,\pm\,$45.21 \\
        \hline
    \end{tabular}
    \caption{Same as \Table{\ref{tab:betas_dP}}, but for a linear regression in the form of Eq.\,(\ref{fit7}) for $T_{\rm eff}$ \cas{(K)}.}
    \label{tab:betas_Teff}
\end{table}

\begin{table}
    \centering
    \begin{tabular}{lcc}
        \hline
        \, & Step overshoot & Exponential overshoot\\
        \hline
        $\beta_0\,({\rm Offset})$ &  3.746$\,\pm\,$0.002 & 3.704$\,\pm\,$0.002 \\
        $\beta_1\,(\alpha_{\rm ov})$ &  -0.2478$\,\pm\,$0.0007 & -3.577$\,\pm\,$0.006\\
        $\beta_2\,(D_{\rm mix})$ & -0.0015$\,\pm\,$0.0001 & -0.000560$\,\pm\,$0.000009 \\
        $\beta_3\,(M_{\star})$ & -0.2440$\,\pm\,$0.0004 & -0.2703$\,\pm\,$0.0003 \\
        $\beta_4\,(Z)$ & -1.22$\,\pm\,$0.02 & -0.61$\,\pm\,$0.01 \\
        $\beta_5\,(X_{\rm ini}) $& 0.506$\,\pm\,$0.003 & 0.543$\,\pm\,$0.003\\
        $\beta_6\,(X_{\rm c}')$ & -0.0080$\,\pm\,$0.0006 & 0.0096$\,\pm\,$0.0005\\
        $\beta_7\,(M_{\rm cc})$ & -3.592$\,\pm\,$0.006 & -2.825$\,\pm\,$0.005 \\
        $\beta_8\,(R_{\rm cc})$ & 10.48$\,\pm\,$0.01 & 10.131$\,\pm\,$0.009 \\
        \hline
    \end{tabular}
    \caption{Same as \Table{\,\ref{tab:betas_dP}}, but for a linear regression
      in the 
form of Eq.\,(\ref{fit7}) for $\log g$ \cas{($g$ given in cm$^{-1}$s$^{-2}$).}}
    \label{tab:betas_log_g}
\end{table}

\cas{Once a
  value for each of the six input parameters is chosen, the mass and radius of the convective core (denoted as $M_{\rm cc}$ and
  $R_{\rm cc}$, respectively) then follow. The computation of $M_{\rm cc}$ and $R_{\rm cc}$ in our setup
  comes through the estimation of the core boundary mixing properties, of which
  the parameters $f_{\rm ov}$ and $D_{\rm mix}$ are simple representations.
  With Eq.\,\ref{fit6}, we represent the $\Pi_0$ from the grid models
  by a multivariate linear statistical estimation, with the aim to have a
  fast tool to compute model properties for parameter sets that fall within the
  range of those of the grids but do not coincide with an actual grid point. The
  fit in Eq.\,\ref{fit6} also highlights the correlations among the six parameters
  that should be kept in mind when trying to fit observed values of $\Pi_0$.  Since the fit
  for $\Pi_0$ from Eq.\,\ref{fit6} is an approximation, there is no longer a
  unique correspondence between the six input parameters and the values for
  $M_{\rm cc}$ and $R_{\rm cc}$ for parameter combinations that do not coincide
  with a grid model. For this reason, we also pay specific attention to
  $M_{\rm cc}$ and $R_{\rm cc}$.}
\begin{figure}
    \centering
    \includegraphics[width = \columnwidth]{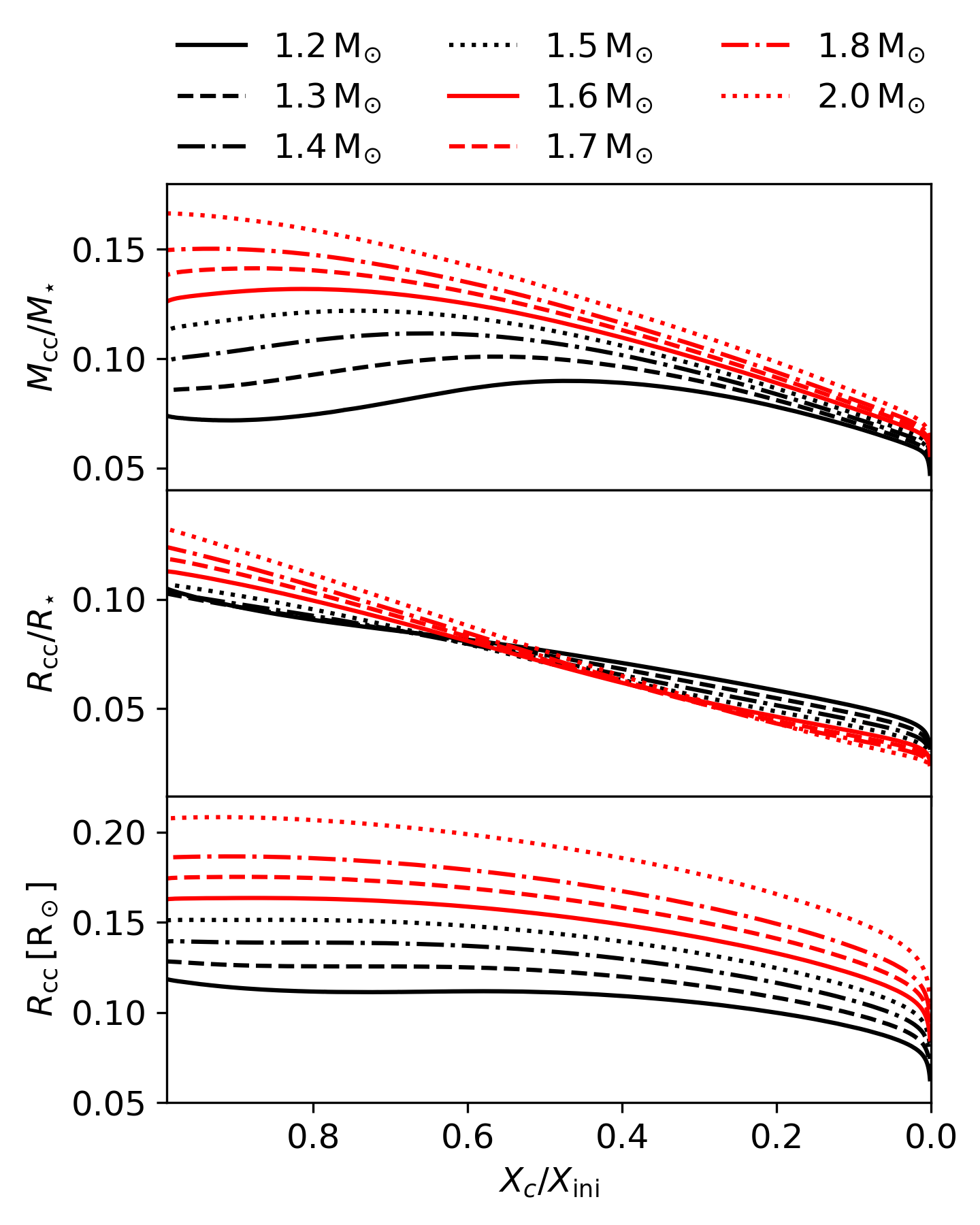}
    \caption{Evolution from near-ZAMS to TAMS of the mass (top panel) and radius (\comm{middle} panel) fraction of the convective core for different stellar masses. \comm{The evolution of the absolute radius of the convective core is given in the bottom panel.} The tracks are for $X_{\rm ini} = 0.71$, $Z = 0.014$, $D_{\rm mix} = \SI{0.1}{\cm\squared\per\second}$ and $f_{\rm ov} = 0.03$. }
    \label{fig:xc-mccrcc}
\end{figure}

\begin{figure}
    \centering
    \includegraphics[width = \columnwidth]{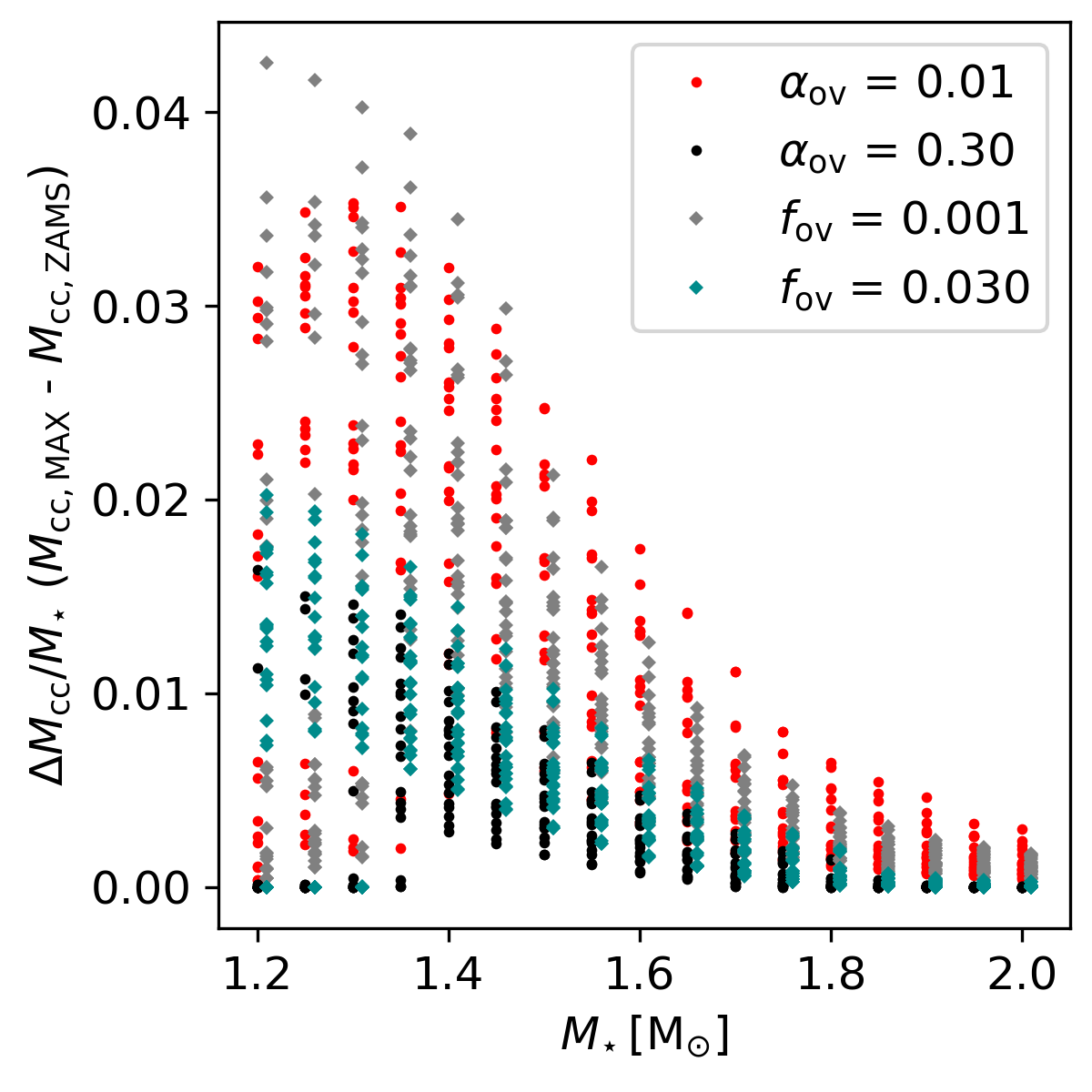}
    \caption{The maximum difference in mass fraction  of the convective core as a
      function of stellar mass for different overshooting prescriptions. 
      For visibility purposes, the data points corresponding to exponential overshoot have been shifted by 0.01\Msun along the \comm{abscissa} axis.}
    \label{fig:M-Mcc}
\end{figure}

Some of the models in the grids have a growing convective core as the
hydrogen-burning progresses, while others have a shrinking convective core,
where the transition occurs roughly around a birth mass of 1.6\Msun \citep[cf.,
Fig.\,3.6 in][]{aerts2010}. This is illustrated for our grids of models in
Fig.\,\ref{fig:xc-mccrcc} in terms of the mass and size of the convective
core. The phenomenon of a shrinking or growing core lies at the basis of the
different correlation structure seen \comm{in the different morphologies of the
  trends for different masses in} Fig.\,\ref{fig:dP-Xc} in terms of the stellar
mass, because the mode cavities are influenced by it during the evolution of the
star. Moreover, the extent of the overshoot zone \comm{affects} the mode
trapping \comm{and shrinkage or growth of the convective core is influenced by
  the core overshooting.}  A summary representation of the maximal growth of the
convective core mass during the hydrogen burning, for the entire mass range of
the grids and for the two descriptions of the overshooting, is \comm{shown} in
Fig.\,\ref{fig:M-Mcc}. The gain in core mass gradually decreases as the birth
mass increases. It disappears for the most massive stellar models in our grid,
as their convective core never grows, but only shrinks after
birth. \cas{Moreover, the evolution of the core mass is strongly dependent on
  the value of the overshoot parameter.}

The results in Figs\,\ref{fig:xc-mccrcc} and \ref{fig:M-Mcc} imply that 
the mass and size of the convective core 
\cas{correlate with $\Pi_0$ and that the correlation structure is different for step
and exponential overshoot.  In order to investigate these dependencies,
we add these two quantities to the linear regression model},
\begin{equation}
    \begin{split}
    \Pi_0  = & \beta_0 + \beta_1 \alpha_{\rm ov} + \beta_2 D_{\rm mix} +
    \beta_3 M_{\star} + \beta_4 Z + \beta_5 X_{\rm ini} + \beta_6  X_{\rm c}' \\ 
    & + \beta_7 M_{\rm cc}' + \beta_8 R_{\rm cc}' + \epsilon,
    \label{fit7}
    \end{split}
\end{equation}
where $M_{\rm cc}'$ is expressed in terms of $M_{\star}$ and $R_{\rm cc}'$ is
expressed in terms of $R_{\star}$.  The core boundary is defined by the Ledoux
criterion and we defined
the core radius $R_{\rm cc}$ as the point
where the Brunt-V\"ais\"al\"a frequency becomes larger than a small threshold
$N^2 > 10^{-9}$rad$^2$s$^{-2}$. 
\cas{ With this set up, we test how a multivariate linear model captures the influence of the core mass and size on $\Pi_0$.} 
The results are listed in
Table\,\ref{tab:betas_dP} \comm{for both grids}.

\cas{ We evaluate the capacity of the multivariate linear model in
  Eq.\,(\ref{fit7}) by inspecting the square-root of the residual sum of squares
  of the fit, averaged over all grid points, denoted here as
  $\sqrt{\langle \rm RSS \rangle}$. We find a value of \SI{252}{\second} for
  step overshoot and \SI{253}{\second} for exponential overshoot.  As these
  $\sqrt{\langle \rm RSS \rangle}$ values are comparable with typical
  measurement uncertainties for $\Pi_0$ (Table\,\ref{tab:MLE_Grid}), we conclude
  that $\Pi_0$ can generally be well approximated by a multivariate linear
  statistical model up to the level of the measurement uncertainties. We do keep
  in mind that the approximation works less well near the TAMS, where
  non-linearity of $\Pi_0$ is larger than for the other evolutionary phases
  (cf., Fig.\,\ref{fig:dP-Xc}).  Non-linear multivariate regression to
  approximate $\Pi_0$ (and $T_{\rm eff}$ and $\log g$ further on) 
is beyond the scope of this paper, but may be interesting
  to consider as an improvement of our current work for future stellar modelling
  of near-TAMS pulsators.}

Both overshooting prescriptions suggest that the amount of mixing in the
radiative zone, $D_{\rm mix}$, has \comm{only a} modest effect on $\Pi_0$, as
already reflected in Fig.\,\ref{fig:dP-Xc} (see also table 2 in
\citealt{johnston2019}).  However, we expect $D_{\rm mix}$ to have some effect on
$\Pi_0$ \citep{VanReeth2015a} and mainly on the mode trapping, as was shown from
forward modelling of \gmodes in the {\it Kepler\/} B-type pulsator KIC\,10526294
\citep{moravveji2015}.  For this reason, we applied the principle of `backward
selection', where one eliminates one-by-one the least significant
$\beta$-parameter and test if the simpler statistical model is more appropriate
\citep[cf.,][where this is explained in more detail and was applied to a similar
problem]{aerts2014}.

A formal way to deduce which statistical model is the more appropriate one is
the Bayesian Information Criterion (BIC). Among many other statistical tests,
the BIC \comm{corrects} for the complexity of a statistical model by applying a
penalty involving the degrees of freedom \citep{claeskens2008} rather
than just using the RSS for the model selection.  Here, we make an application,
using
\begin{equation}
    {\rm BIC} \equiv q \ln\left(\frac{\rm RSS}{q}\right) + k \ln(q),
\end{equation}
where $k$ is the number of free parameters and $q$ is the number of grid points.
Starting with a statistical model described by Eq. (\ref{fit7}), we compute the corresponding $\beta$ \comm{values} and their $p$-value, which is defined in the case $q \gg k$ as,
\begin{equation}
    p = 2[1 - \Phi(|t|)],
\end{equation}
where \comm{$\Phi(|t|)$} is the cumulative standard normal distribution, i.e., the integral of the standard
normal density between $-\infty$ and \comm{$|t|$}. Its argument $t$ is inverse of the relative uncertainty \comm{on \bm{$\beta$}}, where the absolute uncertainties are computed by taking the square-root of the diagonal elements of the \comm {variance-covariance} matrix
\begin{equation}
    \bm{V}(\bm{\beta}) = (\bm{X}^\top \bm{X})^{-1} \sigma^2,
\end{equation}
\comm{in which}
\begin{equation}
    \sigma^2 = \frac{1}{q-(k+1)} (\bm{Y} - \bm{X\beta})^\top(\bm{Y} - \bm{X\beta}), 
\end{equation}
where for our grids $q \gg k$.
Next, we compute a new statistical model where the $\beta_i$ with the highest $p$-value from the previous model is omitted, and the new BIC is evaluated. This process is repeated until the BIC of the new model is higher than the previous one. This previous model is then selected to be the optimal statistical model. \comm{We refer the interested reader to \cite{aerts2018} for more details on the statistical framework discussed here. For the statistical models for $\Pi_0$, none of the parameters can be omitted according to the BIC criterion.}

\subsection{Linear statistical model for \texorpdfstring{$T_{\rm eff}$}{Teff} and \texorpdfstring{$\log g$}{logg}}

\begin{figure*}
    \centering
    \includegraphics[width = \textwidth]{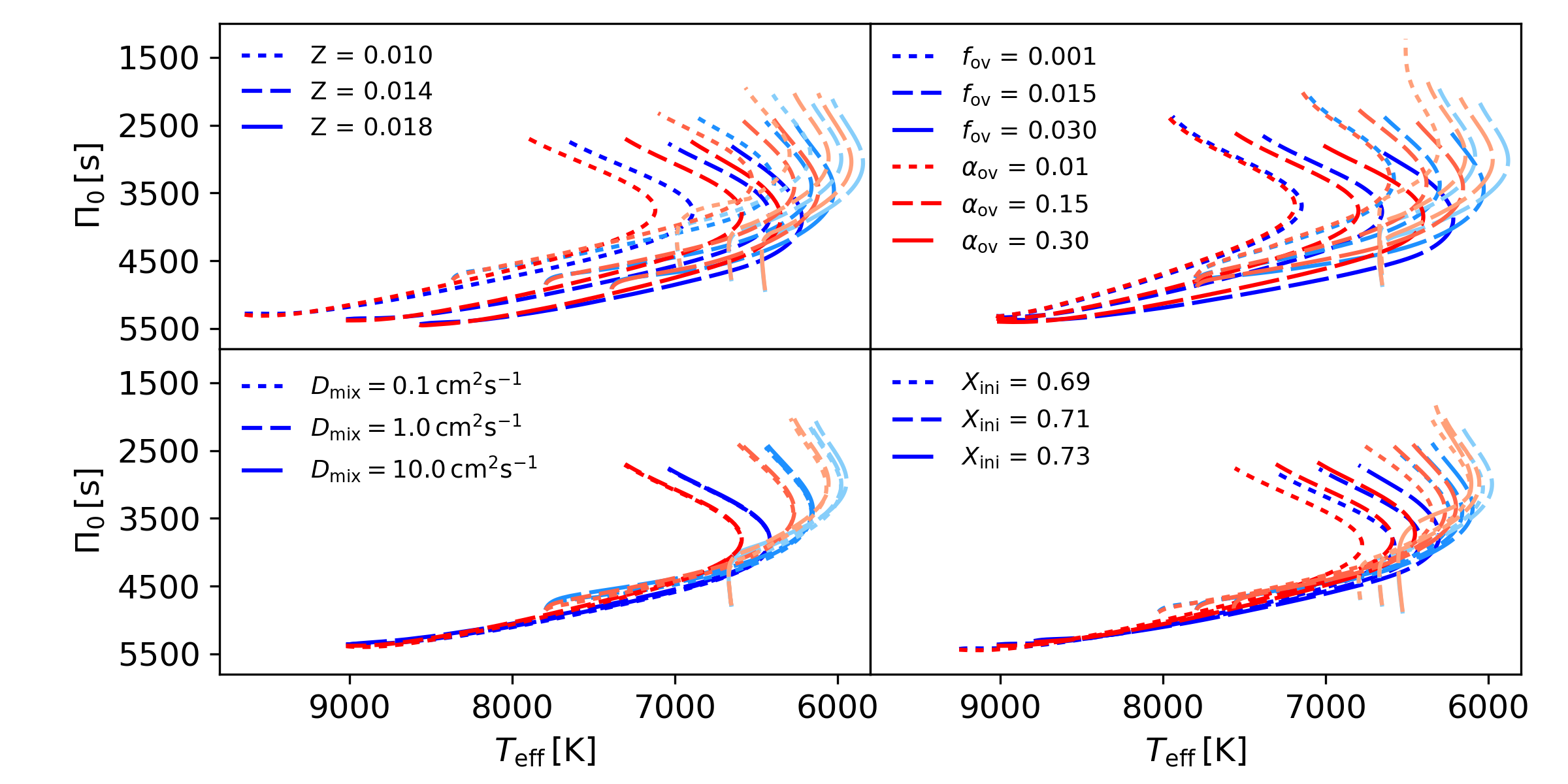}
    \caption{\cas{Correlation between the effective temperature $T_{\rm eff}$
        and the seismic parameter $\Pi_0$ along the main sequence (starting at $X_{\rm c}' = 0.99$) for
        changing metallicity (top left), amount of core overshooting (top right), \comm{the amount} of mixing in the radiative zone (bottom left) and initial hydrogen mass fraction (bottom right), for a 1.3-
        (lightest colours), 1.6- and 1.9-\Msun (darkest colours) star.} The
      blue lines correspond to an exponential overshooting prescription and the
      red lines to step overshooting prescription, both of which consider the
      radiative temperature gradient in the overshoot zone. When a parameter is
      not being varied, it is set at the values $X_{\rm ini}=0.71$, $Z = 0.014$,
      $D_{\rm mix} = \SI{1.0}{\cm\squared\per\second}$,
      $\alpha_{\rm ov} = 0.225$ and $f_{\rm ov} = 0.0225$. Both axis have been
      inverted as to emulate an HR diagram, where the main-sequence stars evolve
      upwards. The rapid change in $\Pi_0$ at constant $T_{\rm eff}$ seen in the
      tracks of the 1.3\Msun models results from the transition from pre-MS to
      the MS. These models are already on the main-sequence track and are burning hydrogen
      in their convective cores. }
    \label{fig:Teff-dP}
\end{figure*}

For 37 of the \gDor stars with identified dipole modes in the sample from
\citet{VanReeth2016}, a measurement of $T_{\rm eff}$, $\log\,g$, and [M/H] has
been obtained from high-resolution spectroscopy. Uncertainties for [M/H] are large ($\sim$100\percent) \citep[fig.\,2 and table~5 in][]{VanReeth2015b}, but the
spectroscopic $T_{\rm eff}$ and $\log\,g$ measurements have relative average precisions of $\sim$2\percent and $\sim$9\percent, respectively, for the stars in our sample.
As the relative uncertainties of $T_{\rm eff}$ and $\log g$ are comparable with the average precision on $\Pi_0$ (for the stars that have spectroscopic data) \comm{these observables are good candidates to lift part of the degeneracy between $\Pi_0$ and the stellar parameters. }
For the statistical models of $T_{\rm eff}$, the values of $\sqrt{\langle \rm RSS \rangle}$ are 233 and \SI{229}{\K} for step and exponential overshoot, respectively. For $\log g$, the statistical model has $\sqrt{\langle \rm RSS \rangle} = 0.05$\,dex for both overshooting prescriptions. For $T_{\rm eff}$ this is larger than the typical uncertainty on the observed \comm{value measured} by \cite{VanReeth2015b}, while for $\log g$ it is significantly better than the observed uncertainties.
As shown in Fig.\,\ref{fig:Teff-dP}, there is a correlation between $\Pi_0$ and $T_{\rm eff}$ from near-ZAMS ($X_{\rm c}' \sim 0.99$) to near-TAMS ($X_{\rm c}' \sim 0$). Again, we vary either $X_{\rm ini}$, $Z$, $\alpha_{\rm ov}/f_{\rm ov}$ or $D_{\rm mix}$ while keeping the other parameters fixed and do this for 1.3, 1.6 and 1.9\Msun. 

\comm{Analogous to} the procedure developed for $\Pi_0$ above, we searched for an
optimal linear multivariate regression model for $T_{\rm eff}$ and $\log g$. Again, we start with a statistical model as described in Eq.\,(\ref{fit7}) and perform backward selection to see if any redundant parameters can be eliminated, by minimizing the BIC. 
For exponential overshoot we find that $D_{\rm mix}$ has little effect on $T_{\rm eff}$, but BIC increases when this parameter is not taken into account, suggesting it is not a redundant parameter. In the case of step overshoot, a decrease in BIC of suggests the model without $D_{\rm mix}$ is statistically favoured. The results of these linear regressions for $T_{\rm eff}$ are listed in \Table{\ref{tab:betas_Teff}}. For the statistical models of $\log g$, we find that none of the parameters may be omitted according to the BIC \comm{(\Table{\ref{tab:betas_log_g}})}. 

\section{Comparing the two overshooting prescriptions}
\label{sec:BM_overshoot}
\comm{A major uncertainty in evolution theory of star born with a convective core} is the efficiency of the mixing inside the core
overshoot region. This translates into the question of the functional
prescription of the overshooting. It was recently shown that \gmodes have the
potential to unravel the most appropriate shape, or at least to discriminate
between an exponential and a step \comm{overshooting prescription \citep{pedersen2018}}. 

Here, we compare theoretical models with step versus exponential overshooting
prescriptions for the mass range of the \gDor stars to test if we can make a distinction \comm{between} our two grids of models.  We do this by
applying the method of parameter estimation and model selection described in
\citet[][problem 2]{aerts2018}. For both overshooting prescriptions, we
chose a benchmark model from one grid and fit \cas{the corresponding $\Pi_0$},
effective temperature and surface gravity to the models in the other grid. \comm{Adopting the notations in
\citet{aerts2018}, we use the subscript to indicate the
`observational' values (BM) and a superscript
to distinguish between a step (s) and an exponential (e) overshooting prescription.} We then compute the optimal \comm{(independent)} parameters 
\begin{equation}
    \bm{\theta}_0 = (M_\star, X_{\rm c}', \alpha_{\rm ov}/f_{\rm ov}, Z, X_{\rm ini}, D_{\rm mix} ),
\end{equation}
for both benchmark models according to

\begin{equation}
\bm{\theta}_0^{(\rm e)} = \arg\min_{i=1}^{r_{\rm e}'}\left[ \left( \bm{Y}^{\rm (e)}_{i} - \bm{Y}^{\rm (s)}_{\rm BM}\right)^\top \left(\bm{\hat{V}}^{\rm (e)}\right)^{-1} \left( \bm{Y}^{\rm (e)}_{i} - \bm{Y}^{\rm (s)}_{\rm BM}\right)\right], 
\end{equation}
\begin{equation}
\bm{\theta}_0^{(\rm s)} = \arg\min_{j=1}^{r_{\rm s}'}\left[ \left( \bm{Y}^{\rm (s)}_{j} - \bm{Y}^{\rm (e)}_{\rm BM}\right)^\top \left(\bm{\hat{V}}^{\rm (s)}\right)^{-1} \left( \bm{Y}^{\rm (s)}_{j} - \bm{Y}^{\rm (e)}_{\rm BM}\right)\right],
\end{equation}
where the \cas{index $i = 1,\ldots ,\num{819774}$ runs over all stellar models in the grid with the exponential overshooting prescription and the index $j = 1, \ldots, \num{1300590}$ runs over all models in the grid with the step overshooting prescription \comm{(superscript on $\bm{\theta_0}$ indicates the corresponding grid)} and}
\begin{equation}
\bm{Y}_{i/j} = 
 \begin{pmatrix}
  \Pi_{0,i/j} \\
  T_{{\rm eff},i/j}   \\
  \log g_{i/j} \\
  \end{pmatrix}.
\end{equation}
Moreover, we define the matrix
\begin{equation}
    \bm{V}(\bm{Y}) = \frac{1}{q'-1}\sum_{k=1}^{q'} \left(\bm{Y}_k - \bar{\bm{Y}} \right)\left(\bm{Y}_k - \bar{\bm{Y}} \right)^\top, 
    \label{eq:V-matrix}
\end{equation}
taking into account the variance across the grid, \comm{where \bm{$\bar{Y}$} is the mean value of $\bm{Y}$ in the grid}.
We choose eight benchmark \cas{\gDor} models; a young low-mass star, an old low-mass star, a young high-mass star and an old high-mass star, with $Z = 0.014$, $X_{\rm ini} = 0.71$, $\alpha_{\rm ov}(f_{\rm ov}) = 0.15(0.015)$ and $D_{\rm mix} = \SI{1.0}{\cm\squared\per\second}$, \comm{all} for both overshooting prescriptions. In \Table{\ref{tab:bm}} we list the maximum likelihood estimate (MLE) from using an incorrect overshooting prescription for each of these eight benchmark models. For all of these models, the resulting $\bm{\theta}_0$ from fitting $(\Pi_0, T_{\rm eff}, \log g)$ does not change drastically, compared to the step size of the grids for $M_\star$ and $X_{\rm c}'$. The grids contain too few points in the other parameters to draw a conclusion on how robust the MLE is for these parameters. However, even though there is scatter in $\alpha_{\rm ov}/f_{\rm ov}$, $D_{\rm mix}$, $Z$ and $X_{\rm ini}$, similar masses and age-proxies are found.

\begin{figure}
    \centering
    \includegraphics[width = \columnwidth]{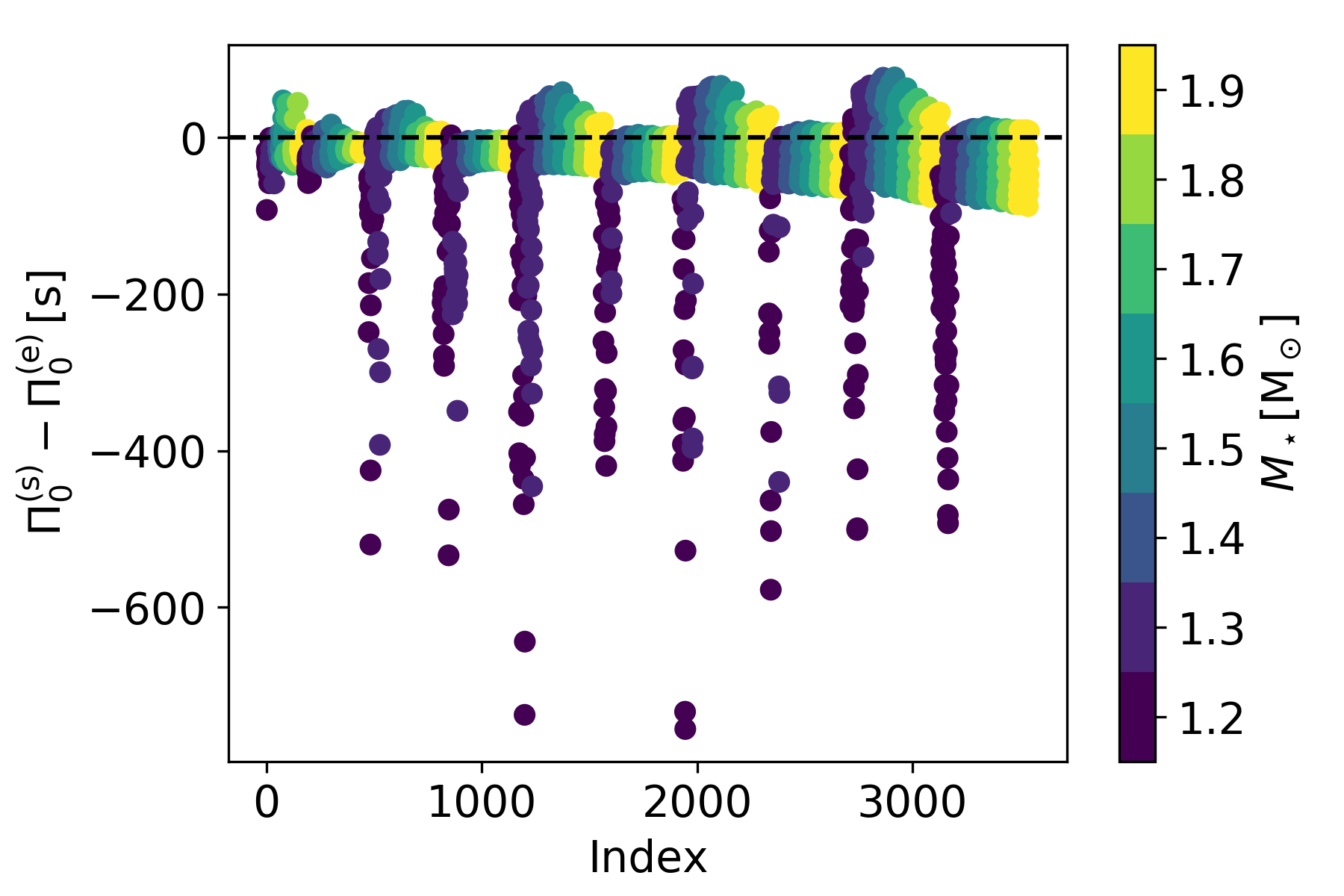}
    \caption{Difference \comm{between} $\Pi_0$ in the step overshoot grid
      ($\Pi_0^{\rm (s)}$) and the exponential overshoot grid
      ($\Pi_0^{\rm (e)}$), for pairs of models with the same parameters
      $M_\star$, $X_{\rm c}'$, $D_{\rm mix}$, $Z$, $X_{\rm ini}$, \comm{varied across
        the grid.} An $\alpha_{\rm ov}$ in one grid corresponds to
      $10f_{\rm ov}$ in the other grid. \cas{The pairs of models are equally
        distributed across both grids and are arbitrarily labelled from 1 to
        3541.} When $M_\star \geq 1.4$\Msun the maximum difference is comparable
      with a typical observational uncertainty on $\Pi_0$. For visual aid, a
      dashed line is plotted at zero difference.}
    \label{fig:compare_overshoot_dP}
\end{figure}

Furthermore, we note that for $\alpha_{\rm ov} = 10 f_{\rm ov}$, the MLE agrees
\comm{well} with the input parameters (see
Fig. \ref{fig:compare_overshoot_dP}). This scaling with a factor 10 between
$\alpha_{\rm ov}$ and $f_{\rm ov}$ is a commonly used rule-of-thumb when
comparing step overshoot and exponential overshoot. In
Fig.\,\ref{fig:compare_overshoot_dP}, the difference in $\Pi_0$ is plotted
between two models, each from a different grid, that have the same parameters
($X_{\rm c}'$ within 0.001) where $\alpha_{\rm ov} = 10 f_{\rm ov}$.  This scaling
rule does give similar values for \cas{ $\Pi_0$} within the parameter space of
\gDor stars, compared to typical uncertainties on this seismic parameter, as the
difference is less than \SI{60}{\second} for masses above 1.4\Msun. The
deviation between the two grids \comm{at the lower masses is a result from the
  fact that this factor 10 is not exact and also mass dependent} \citep[11.36
$\pm$ 0.22 was found by][]{claret2017}.  We stress that our inability to
distinguish between step and exponential overshooting from the MLE is due to the
coarseness of the grids \comm{(\Table{\ref{tab:grid}})}, as this determines how
well $\bm{Y}^{\rm (obs)}$ can be matched in the other grid. This can be seen in
the last three columns of \Table{\ref{tab:bm}}.

\begin{table*}
    \centering
    \begin{tabular}{ccccccccccc}
         \hline
         & $M_\star\,$[\Msun] & $X_{\rm c}'$ & $\alpha_{\rm ov}$ & $f_{\rm ov}$ & $D_{\rm mix}\,$[cm$^{-2}$s$^{-1}$] & $Z$ & $X_{\rm ini}$ & $ \Pi_0\,$[s] & $T_{\rm eff}\,$[K] & $\log g$ \\
        \hline

        BM & 1.30 &  0.90 & 0.150 & - & 1.0  &  0.014 & 0.71 & 4327  &  6669 & 4.33 \\
        GRID & 1.35 &  0.92 & - & 0.0075 &  10.0  &  0.014 & 0.73 & 4330  &  6678 & 4.33 \\
        \hline
        BM & 1.30 &  0.90 & - & 0.0150 &  1.0  &  0.014 & 0.71 & 4352  &  6669 & 4.33 \\
        GRID & 1.25 &  0.88 & 0.225 & - & 1.0  &  0.014 & 0.69 & 4343  &  6657 & 4.33 \\
        \hline
        BM & 1.30 &  0.10 & 0.150 & - & 1.0  &  0.014 & 0.71 & 3140  &  6158 & 4.04 \\
        GRID & 1.35 &  0.10 & - & 0.0010 &  10.0  &  0.018 & 0.69 & 3147  &  6186 & 4.04 \\
        \hline
        BM & 1.30 &  0.10 & - & 0.0150 &  1.0  &  0.014 & 0.71 & 3182  &  6104 & 4.01 \\
        GRID & 1.25 &  0.10 & 0.225 & - & 1.0  &  0.014 & 0.69 & 3179  &  6099 & 4.02 \\
        \hline
        BM & 2.00 &  0.90 & 0.150 & - & 1.0  &  0.014 & 0.71 & 5551  &  9217 & 4.29 \\
        GRID & 2.00 &  0.90 & - & 0.0150 &  0.1  &  0.014 & 0.71 & 5535  &  9216 & 4.29 \\
        \hline
        BM & 2.00 &  0.90 & - & 0.0150 &  1.0  &  0.014 & 0.71 & 5531  &  9218 & 4.29 \\
        GRID & 2.00 &  0.90 & 0.150 & - & 1.0  &  0.014 & 0.71 & 5551  &  9217 & 4.29 \\
        \hline
        BM & 2.00 &  0.10 & 0.150 & - & 1.0  &  0.014 & 0.71 & 4172  &  7100 & 3.69 \\
        GRID & 1.80 &  0.17 & - & 0.0225 &  10.0  &  0.010 & 0.71 & 4166  &  7088 & 3.69 \\
        \hline
        BM & 2.00 &  0.10 & - & 0.0150 &  1.0  &  0.014 & 0.71 & 4210  &  6957 & 3.64 \\
        GRID & 1.90 &  0.14 & 0.300 & - & 1.0  &  0.010 & 0.73 & 4221  &  6979 & 3.63 \\
       \hline
    \end{tabular}
    \caption{The MLEs for eight different benchmark models (BM) using a grid with the incorrect overshooting prescription, i.e., step overshoot models are fitted to an exponential overshoot grid and vice versa.}
    \label{tab:bm}
\end{table*}

\section{Parameter estimation from 
\texorpdfstring{$\Pi_0$}{Pi0}, \texorpdfstring{$T_{\rm \lowercase{eff}}$}{Teff}, and \texorpdfstring{$\log \lowercase{g}$}{logg}}
\label{sec:Mahalanobis}

\begin{figure*}
    \centering
    \includegraphics[width = \textwidth]{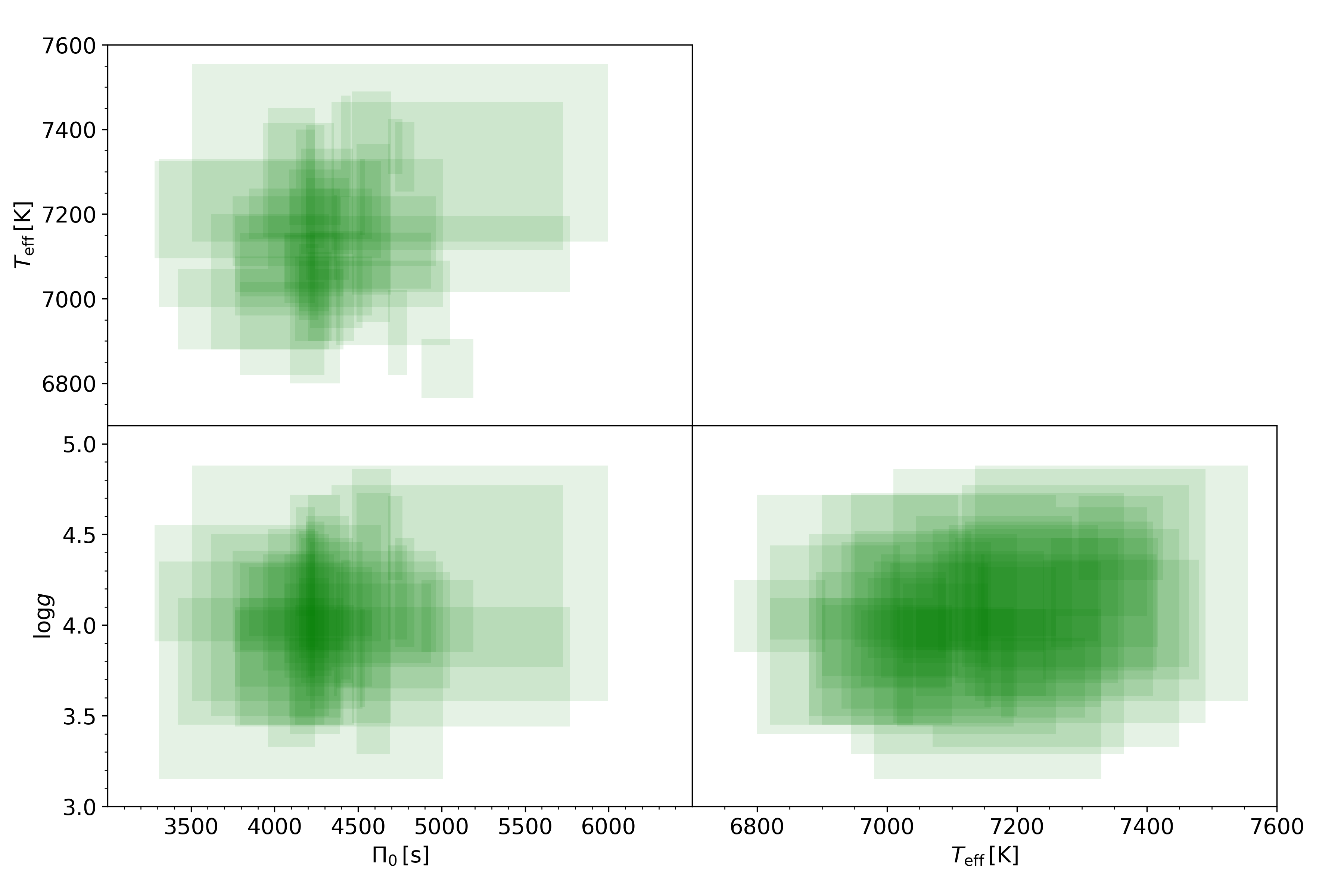}
    \caption{\cas{2D projected 1$\sigma$ error boxes of the observable ($\Pi_0$, $T_{\rm eff}$, $\log g$) for the 37 stars in our sample for which spectroscopic measurements are available.}}
    \label{fig:1sigma_obs}
\end{figure*}

We aim to test the feasibility of estimating (some of) the six parameters in the
two grids for all \gDor stars in the sample of \citet{VanReeth2016} with $\Pi_0$
deduced from prograde dipole modes in
an automated way. We recall that these measured values of $\Pi_0$ were
derived along with estimation of the near-core rotation frequency
\citep{VanReeth2016,VanReeth2018}. Our aim is to test if $\Pi_0$, along with spectroscopic measurements of $T_{\rm eff}$ and $\log g$
are sufficient to estimate $M_{\star}$, $ X_{\rm c}'$, $\alpha_{\rm ov}/f_{\rm ov}$, and possibly $Z$,
$X_{\rm ini}$ and/or $D_{\rm mix}$ from an ensemble of \gDor stars (see Fig. \ref{fig:1sigma_obs}, \comm{illustrating the 2D projected 1$\sigma$ errors on the observed parameters of each star}). We use an automated grid-based
approach, keeping in mind the uncertainties of individual theoretically computed
g-mode frequencies due to unknown aspects of the input physics of the stellar
models, as assessed by \citet{aerts2018}. These authors describe a new method for forward seismic modelling, where the correlation between the \comm{observables} is taken into account. This method is based on the Mahalanobis distance which is defined in our 3D case as
\begin{equation}
    D_{{\rm M},j} = \left(\bm{Y}_j^{(\rm th)} - \bm{Y}^{(\rm obs)}\right)^\top \left(\bm{\Lambda} + \bm{V}\right)^{-1} \left(\bm{Y}_j^{(\rm th)} - \bm{Y}^{(\rm obs)}\right).
\end{equation}
The matrix $\bm{\Lambda}$ = diag$(\sigma_{\Pi_0}^2, \sigma_{T_{\rm eff}}^2, \sigma_{\log g}^2)$ takes the uncertainties of the observed quantities into account and the (co-)variance matrix $\bm{V}$ is computed according Eq.\,(\ref{eq:V-matrix}) for both grids. \newline
\newline
We first get MLEs for all stars in our sample for which spectroscopic data is
available by minimizing the Mahalanobis distance across \comm{each of} the
grids. As has already been demonstrated with the statistical models presented in
Section~\ref{sec:stat_mod}, some parameters are \comm{more strongly} correlated
with the observables than others. We therefore assess whether it is possible to
reduce the dimension of our forward modelling problem from a principal component
analysis (PCA). For each star, the best 10\percent of the models is selected and
the basis of this 6D solution space is redefined in terms of its principal
components (PCs). These PCs are the eigenvectors of the correlation matrix and
the corresponding normalised eigenvalues give the percentages of the total
variance that is captured by each PC. As a rule-of-thumb, the number of PCs
\comm{selected} is the number of PCs which is needed to describe at least
80\percent of the total variance \citep{jolliffe2002}. In this way, we find that
four PCs are sufficient for all the 37 stars in the sample. 

\cas{ The PCs that can be dropped are predominantly connected with
    $D_{\rm mix}$ and $X_{\rm ini}$.  Therefore, we wish to remove these two
    parameters from the grid to reduce the dimensionality, by fixing a value for
    them. In order to assign the most appropriate fixed value, we consider each
    of the six combinations of these two parameters available in the grids and
    determine the MLE for all stars in our sample.} For the four estimated
  parameters, the average standard deviations between these six MLEs is found to
  be $\langle \sigma_{M_\star}\rangle = 0.06$\Msun,
  $\langle \sigma_{X_{\rm c}}\rangle = 0.02$,
  $\langle \sigma_{\alpha_{\rm ov}}\rangle = 0.061$ and
  $\langle \sigma_{Z}\rangle = 0.001$, for step overshoot. For exponential
  overshoot, $\langle \sigma_{M_\star}\rangle = 0.06$\Msun,
  $\langle \sigma_{X_{\rm c}}\rangle = 0.03$,
  $\langle \sigma_{f_{\rm ov}}\rangle = 0.062$ and
  $\langle \sigma_{Z}\rangle = 0.001$. Based on these values, we argue that
  $D_{\rm mix}$ and $X_{\rm ini}$ can be set to the fixed values
  \SI{1.0}{\cm\squared\per\second} and 0.71, respectively.

The results of the
parameters estimations from MLE are listed in \Table{\ref{tab:MLE_Grid}} for
both overshooting prescriptions. The corresponding fractional mass and
fractional radius of the convective core of the best model are also listed. For
$M_\star$, $X_{\rm c}'$, $M_{\rm cc}'$ and $R_{\rm cc}'$ \comm{(fractional mass and
  radius of the convective core), the grid is comprised of a sufficiently large
  number of points to make a comparison. However, for these parameters there is
  no obvious discrepancy reported between the two overshooting prescriptions for
  any given star.}

\begin{figure*}
    \centering
    \includegraphics[width= 0.98\textwidth]{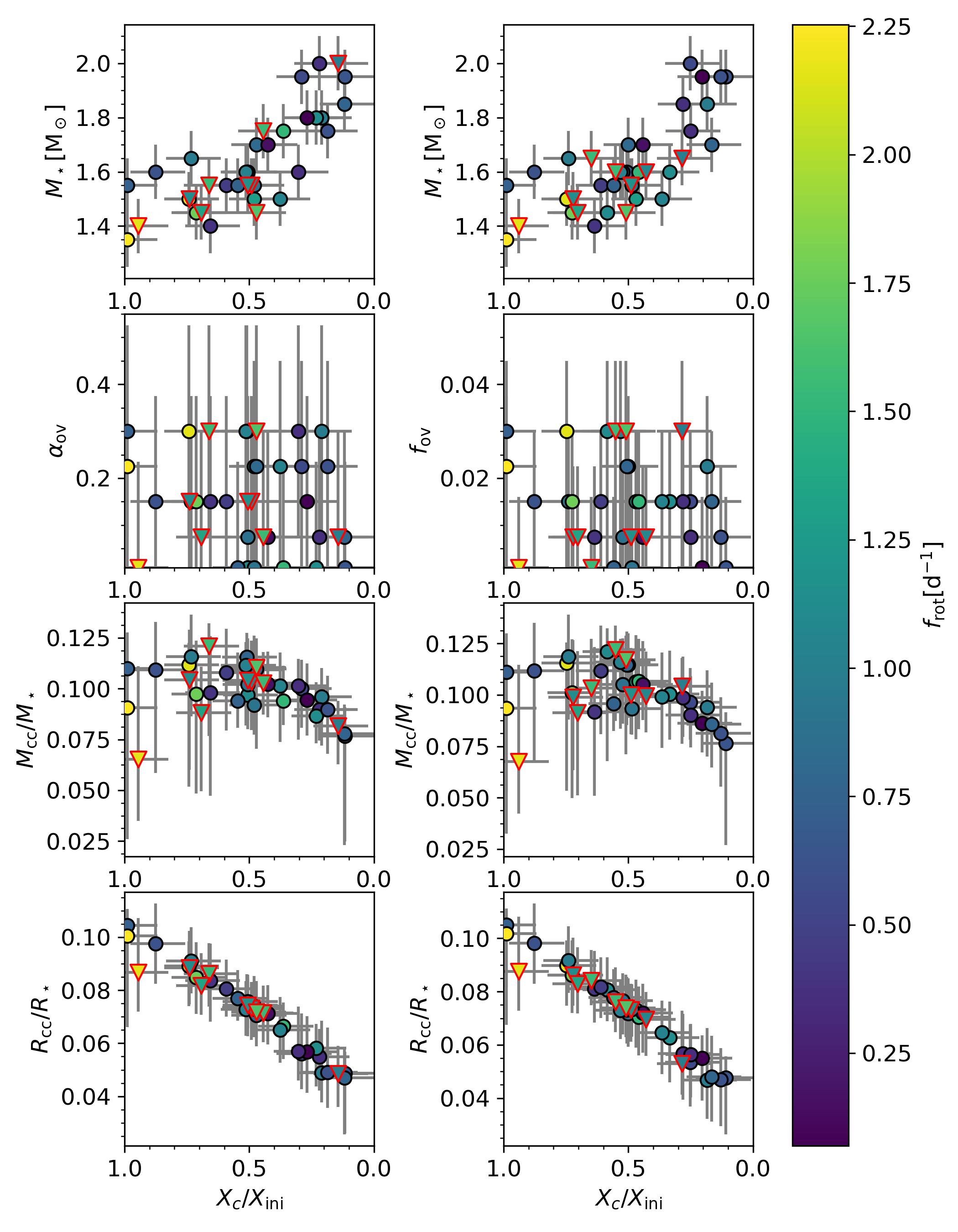}
    \caption{Top to bottom: MLEs for $X_{\rm c}'$ versus mass, overshoot, convective
      core mass and convective core radius, colour-coded by the measured
      near-core rotation frequency from \citet{VanReeth2016}, for all 37 stars
      in our sample. Left column: Step overshoot. Right column: Exponential
      overshoot. Stars that have \comm{observed} Rossby modes in
      addition to gravito-inertial prograde dipole modes are plotted as \cas{triangles}.}
    \label{fig:Xc_f_rot_master}
\end{figure*}

The relation between the near-core rotation frequency $f_{\rm rot}$ and the
evolutionary state of the stars in our sample has \comm{previously} been
investigated by \citet[][their fig.\,1]{aerts2017}, where the spectroscopic
$\log g$ was used as a proxy for the evolutionary stage. In
Fig.\,\ref{fig:Xc_f_rot_master}, $\log g$ is replaced by the MLE of $X_{\rm c}'$ for
both overshooting prescriptions. We colour-code for the measured near-core
rotation rate, for which the relative uncertainty is typically a few
per~cent. From these \comm{more precise} estimates of the \comm{proxy for
  stellar age}, the data reveal that the \comm{cores of intermediate-mass A and
  F stars spin down} as the stars evolve. 

\citet{ouazzani2018} report that faster rotators are less massive and younger
than the slow rotators. This is consistent with our findings that the two
aforementioned faster rotating near-ZAMS stars are less massive than the slower
rotating stars with roughly the same $X_{\rm c}'$, as shown in the top row of
Fig.\,\ref{fig:Xc_f_rot_master}. As the \gDor instability region is
\comm{relatively} small \citep{bouabid2013}, \comm{it is possible for stars to
  be born on the ZAMS outside of the instability region and later evolve into
  it}, and vice versa. \comm{Therefore, it is expected that high-mass \cas{\gDor}
  stars are slow rotators since they are closer to the TAMS than low-mass
  stars.} Yet, the fact we do not observe any evolved fast rotators might also
be caused by an observational selection bias, \comm{because the structure of a
  TAMS star combined with fast rotation creates a dense, and typically
  unresolvable spectrum of pulsation modes (e.g., \citealt{buysschaert2018}).}
\cas{Nevertheless, the fact that the fastest rotators in our sample are all
  found to be relatively young suggests that the transport of angular
  momentum from the core to the envelope must be efficient already early on in
  the stellar evolution. Given that the convective core size decreases as the star evolves (cf., Fig.\,\ref{fig:xc-mccrcc}), our findings suggest that the efficiency of angular momentum transport is somehow connected to the convective core, as already hinted at by \citet{aerts2019}.} \car{So far, the individual
roles of gravito-inertial and Rossby modes in angular momentum transport remain unclear. As can been seen in Fig.\,\ref{fig:Xc_f_rot_master}, we find stars with observed Rossby modes (plotted as triangles) to be situated across the main sequence, except close to the TAMS. }

In the second row of Fig.\,\ref{fig:Xc_f_rot_master} we plot the MLEs of the
overshoot versus $X_{\rm c}'$ and colour-code by the rotation rate to investigate if a
higher overshooting estimate \cas{corresponds with faster rotation at the
  interface of the convective core and the radiative envelope. For both
  overshooting prescriptions, the MLE of the overshooting shows no correlation
  between the amount of core overshooting mixing
  and the near-core
  rotation frequency. This result is complementary to the conclusion by
  \citet{aerts2014} that the surface nitrogen excess of a sample of OB stars,
  which demands efficient mixing in the stellar envelope, is not correlated with
  their surface rotation frequency but rather with the frequency of the pressure
  mode with dominant amplitude. 
  Taking the results of both studies together suggest that core boundary mixing may have various causes, such as rotational mixing \citep[e.g.,][]{brott2011} or pulsational mixing induced by waves \citep[e.g.,][]{RogersMcElwaine2017}, but that its shape remains to be determined. In our modelling, these various kinds of mixing were coded with two functional shapes with a free parameter, i.e., as core overshooting and envelope mixing, without specifying the underlying physical phenomenon. The estimated parameters of these two prescriptions, $f_{\rm ov}$ (or $\alpha_{\rm ov}$) and $D_{\rm mix}$, do not reveal what causes the near-core mixing, but only provide a rough level of mixing required to explain the g-mode behaviour. We find that we cannot pinpoint particular values for our ensemble of stars, suggesting that different physical phenomena may lie at the basis of the overall level of near-core mixing. This is in agreement with the findings for the more massive B-type pulsators \citep{aerts2015}.}

In the last two rows of Fig.\,\ref{fig:Xc_f_rot_master} the fractional mass and
radius of the convective core are placed on the \comm{ordinate} axis. \cas{We
  see that estimation of the convective core mass and size is well achieved,
  despite the absence of predictive power for the core overshoot parameter. This
  is the same conclusion as found by \citet{johnston2019} for three
  gravity-mode pulsators in close binaries and points to a strong probing power
  of dipole gravity modes (through $\Pi_0$) to assess convective core
  properties.}  When we compare the four stars closest to ZAMS, we notice that
the two stars with a lower $f_{\rm rot}$ have more massive and larger convective
cores than the two faster rotating stars. \comm{However}, the overlapping
confidence intervals do not allow us to \comm{claim a significant dichotomy with
  certainty}.

\section{Confidence intervals from ensemble modelling} \label{sec:err_est}
Determining uncertainty intervals from the distributions of the Mahalanobis
distance of each individual star (\citealt{aerts2018}; problem 1) \comm{is not
  meaningful} because of the large span of these intervals in the grids. This
should not come as a surprise, given that one is then determining an estimator
based on a single star, i.e., a sample of size one. A convenient way to deal
with this consists of considering the sample of $S = 37$ as an ensemble. This is
taken to mean a collection of stars that, while having star-specific
characteristics and hence parameters $\bftheta_s$, makes up a sufficiently
`natural' family fulfilling the same underlying theory of stellar structure.

Evidently, when a sample of size $S$ is available, the average of the parameters
$\bftheta_s$, say $\bftheta$, can be estimated much more precisely than the
parameter of an individual star. If we think of a given star parameter
$\bftheta_s$ as being decomposed into $\bftheta_s=\bftheta+\bft_s$, with
$\bft_s$ the star-specific deviation around $\bftheta$, then we can use the
precision of $\widehat{\bftheta}$, \comm{the MLE}, as a measure of precision for
the entire sample. Precisely, we write the likelihood for the entire sample as

\begin{equation}
\begin{split}
L= &\prod_{s=1}^S\frac{1}{(2\pi)^{P/2}|V_s|^{-1/2}} \times \\  
&\exp
\left\{
-\frac{1}{2}
\left[\BY_s(\bftheta_s)-\BY_s^\ast\right]^\top
V_s^{-1}
\left[\BY_s(\bftheta_s)-\BY_s^\ast\right]
\right\}, \\
\end{split}
\end{equation}
with $P=\dim(\BY)$. In our application, $S=37$ and $P=3$.
The kernel of the log-likelihood is
\begin{equation}
\ell=-\frac{1}{2}
\sum_{s=1}^S\ln|V_s|-
\frac{1}{2}\sum_{s=1}^S
\left[\BY_s(\bftheta_s)-\BY_s^\ast\right]^\top
V_s^{-1}
\left[\BY_s(\bftheta_s)-\BY_s^\ast\right]
.
\end{equation}
Consider now the deviance function for the $k$th component of $\bftheta$
\begin{eqnarray}
D(h)&=&
-\sum_{s=1}^S
\left[\BY_s(\bftheta_s)-\BY_s^\ast\right]^\top
V_s^{-1}
\left[\BY_s(\bftheta_s)-\BY_s^\ast\right]
\\ \nonumber
&&+
\sum_{s=1}^S
\left[\BY_s(\bftheta^{(k,h)}_s)-\BY_s^\ast\right]^\top
V_s^{-1}
\left[\BY_s(\bftheta^{(k,h)}_s)-\BY_s^\ast\right], 
\end{eqnarray}
with $\bftheta^{(k,h)}_s=\bftheta_s$, except in the $k$th component, where we put
$\theta^{(k,h)}_{s,k}=\theta_{s,k}+h$, \comm{where $h$ is a (discrete) perturbation.}
We then search for both the negative and positive $h$ values, say $h_L$ and $h_U$, that satisfy $D(h_L)=D(h_U)=3.84$ which is the critical value of the $\chi^2_1$ distribution. When making use of a sufficiently fine grid, \comm{those} grid points can be chosen that satisfy the above requirement sufficiently well. The corresponding confidence interval is:
$$[\theta_{s,k}+h_L;\theta_{s,k}+h_U].$$
Because the likelihood ratio is not based on a quadratic approximation, the interval is not necessarily symmetric, but it will be more symmetric in larger samples. In Appendix \ref{AppendixA2}, the deviance function is plotted as a function of the four parameters we have estimated. The grid is quite coarse for overshoot and $Z$ to properly sample $D(h)$, hence we stay on the conservative side and pick the values of $h_L$ and $h_U$ as the smallest discrete values for which the deviance function is larger than 3.84. For $X_{\rm c}'$, the confidence interval may, statistically seen, contain unphysical values and therefore we truncate the interval at the edge of the grid. The confidence interval of the overshoot is only truncated at the lower edge of the grid. \cas{For both overshoot grids we were not able to determine $h_L$ for our ensemble, therefore the lower limit on $\alpha_{\rm ov}$ and on $f_{\rm ov}$ is always the lowest value in the respective grid.} In Appendix \ref{AppendixA2}, the sampling of $D(h)$ is shown for the four parameters that estimated from the the MLE. This yields uncertainties of 0.1\Msun on $M_\star$, 0.004 on $Z$ (in these cases symmetric), \comm{all rounded to the nearest step in the respective grids} and 0.12 and 0.10 on $X_{\rm c}'$ for the lower and upper uncertainties, respectively.
The uncertainties on $M_{\rm cc}'$ and $R_{\rm cc}'$ are defined as the minimum and maximum values found in the models that lay within the confidence intervals of the estimated parameters. 

\section{Discussion \& Conclusions}
\label{sec:conclusions}
In this paper we have explored the power of \cas{ $\Pi_0$ estimated from prograde
  dipole gravito-inertial modes and in some cases also from Rossby modes}, combined with spectroscopic measurements of
the effective temperature and surface gravity to \cas{ perform asteroseismic
  modelling of \gDor stars, which have a well-developed convective core.}  We
\comm{devised} recipes relating the most important fundamental stellar
parameters to these three observables, from multivariate linear regression, \cas{and have shown that these linear recipes have a proper predicting capacity with respect to the measurement uncertainties.
 The recipes are advantageous in the forward modelling of \gDor stars,
  as they imply a huge decrease in computation time while still providing good
  constraints for the stellar parameters in the high-dimensional parameter
  space.} The method of backward selection based on the BIC reveals that, among
the mass, metallicity, central and initial hydrogen mass fraction, mass and size
of the convective core and the \comm{amount} of mixing in the radiative zone,
only the latter can be \cas{ ignored in the linear recipes. Previously it was found
  that} fitting frequencies of individual trapped \cas{ gravity modes} does
require extra mixing in the radiative envelope \citep{moravveji2015,
moravveji2016, schmid2016}.

The forward modelling of \gDor stars is a cumbersome task \cas{ if the aim is to
  achieve stellar parameters with a relative precision of 10\percent or better. This
  can only be reached when $\Pi_0$, $T_{\rm eff}$, and $\log g$ are measured
  with high precision}.  This problem is illustrated in
Fig.\,\ref{fig:1sigma_obs} where the 2D projected 1$\sigma$ error boxes for
$\bm{Y}^{\rm (obs)}$ are plotted for all 37 stars in our sample. Even though we
use high-precision spectroscopy here, the error boxes are still relatively
large \cas{ due to the correlated nature of the three observables.}

\cas{ This work contains the first asteroseismic forward modelling of \gDor stars
  as an ensemble, using the Mahalanobis distance as described in
  \citet{aerts2018}. With this method, we assume the sample stars to adhere to a
  single underlying stellar evolution theory and that each star with its own
  parameters can be seen as a deviation of an average star in the ensemble
  to derive meaningful uncertainties to go along with the MLEs of $M_\star$,
  $X_{\rm c}'$, $\alpha_{\rm ov}/f_{\rm ov}$, $Z$, as well as the mass and radius of
  the convective core for the 37 \gDor stars in our sample.}
  From PCA it was
concluded that the dimensionality of this problem could be reduced to a 4D
problem when fitting $(\Pi_0, T_{\rm eff}, \log g)$, by fixing $X_{\rm ini}$ and
$D_{\rm mix}$.  For $M_\star$ and $X_{\rm c}'$ we find in general consistent results
between a step and an exponential overshooting prescription. 

We have demonstrated that \cas{ linear statistical models are able to capture the
  correlated nature of the three observables $\Pi_0$, $T_{\rm eff}$ and $\log g$, and
  the fundamental stellar parameters, up to the level of the
  measurement uncertainties. This finding allows us to use recipes rather than
  having to compute dense stellar model grids when fitting the
  \comm{observables} $(\Pi_0, T_{\rm eff}, \log g)$ for future applications to
  additional $\gamma\,$Dor stars.}

Future work \comm{will
  involve} the addition of $X_{\rm ini}$, $D_{\rm mix}$ and $\alpha_{\rm MLT}$
in the modelling of the morphology of the period spacing patterns for all
$\gamma\,$Dor stars with suitable mode detection and spectroscopy,
rather than just $\Pi_0$. Especially the mixing profile in the radiative
zone cannot be constrained with the observables in this paper, but does have
\comm{a} large effect on the morphology of the period spacing pattern. The dips
in the pattern caused by mode-trapping decrease when the mixing efficiency
increases as this process washes out the chemical gradient and therefore reduces
the effect of mode trapping, as shown in B-type stars \citep{moravveji2015}. As
computing the individual theoretical frequencies is a more computationally
demanding exercise, the MLEs of $M_\star$, $X_{\rm c}'$,
$\alpha_{\rm ov}/f_{\rm ov}$ and $Z$ presented in this paper provide a starting
point for refined modelling to estimate $D_{\rm mix}(r)$. \comm{The ongoing NASA
  TESS \citep{ricker2015} and upcoming ESA PLATO \citep{rauer2014} space
  missions will deliver many new g-mode pulsators covering a mass range of 1.2
  to $\sim$20\Msun. Therefore, the similar analysis of potentially thousands of
  new g-mode pulsators and its extension to include binary star information
  (e.g., \citealt{johnston2019}) with the next generation of space telescopes
  will provide an improved insight of stellar structure and evolution of stars
  with convective cores across the HR diagram.}

\section*{Acknowledgements}
\addcontentsline{toc}{section}{Acknowledgements} \cas{We thank the referee,
  Prof.~Hiromoto Shibahashi, for his insightful and detailed comments on our
  manuscript, which improved its clarity
  and quality.}  We thank Bill Paxton, Rich Townsend, and their code development
team members for all their efforts put into the public stellar evolution and
pulsation codes MESA and GYRE.  The research leading to these results has
received funding from the European Research Council (ERC) under the European
Union's Horizon 2020 research and innovation programme (grant agreement
N$^{\rm o}$670519: MAMSIE).  We gratefully acknowledge the Th\"uringer
Landessternwarte in Tautenburg, Germany, for the computation time on their
computer cluster.  \comm{TVR gratefully acknowledges support from the Australian
  Research Council, and from the Danish National Research Foundation (Grant
  DNRF106) through its funding for the Stellar Astrophysics Centre (SAC).}


\bibliographystyle{mnras}
\bibliography{main} 

\appendix
\section{Mode cavities in a typical \texorpdfstring{\gDor}{gDor} stellar model}
\comm{Here, we show an example of the mode cavities of the gravity (g)~modes 
and pressure (p)~modes in a \gDor star.}
\label{Appendix_MC}
\begin{figure}
    \centering
    \includegraphics[width = \columnwidth]{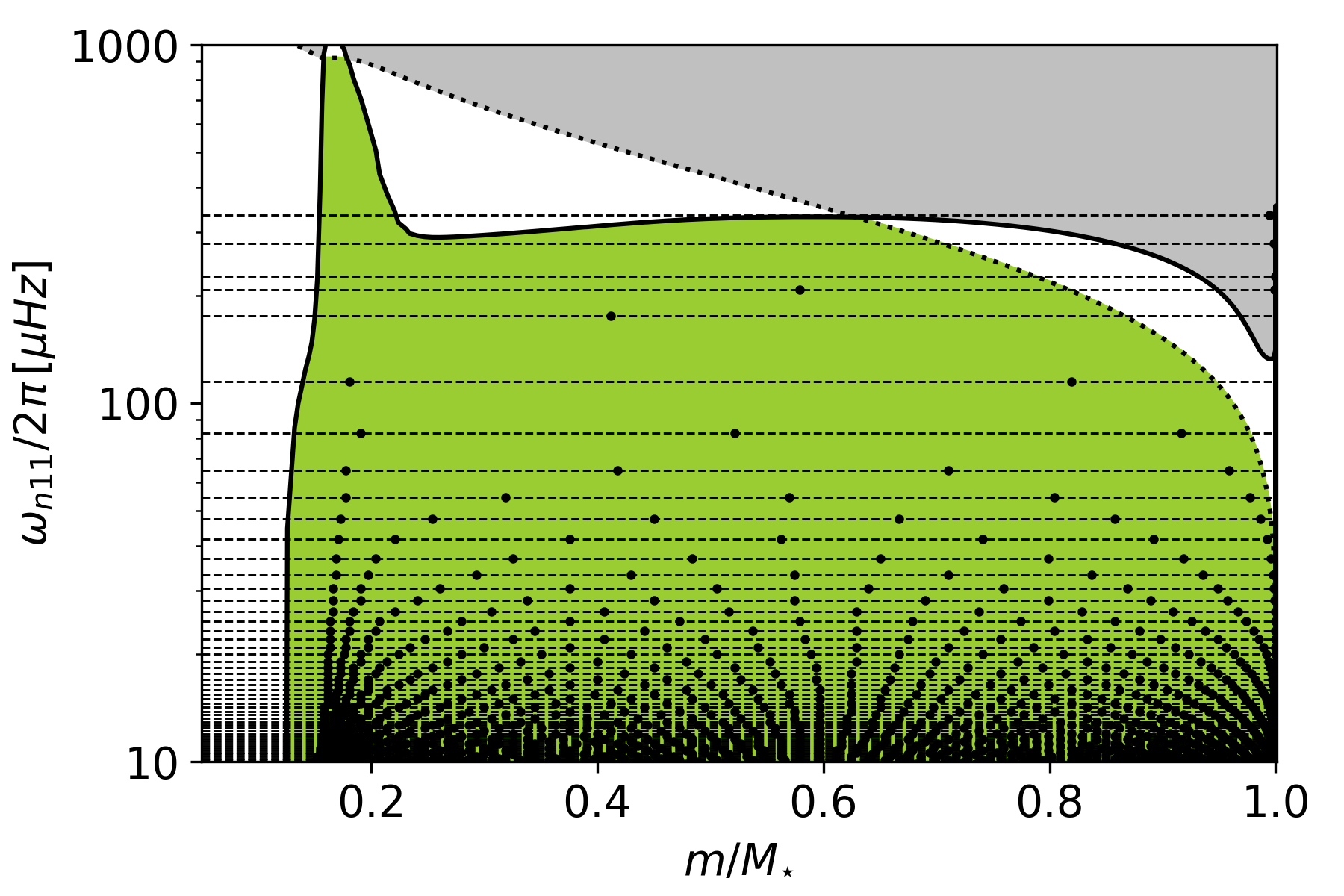}
    \caption{\cas{ Mode cavities of the \gmodes (green, $n_{\rm pg}\,\in\,[-50, -1]$) and p~modes (grey, $n_{\rm pg}\,\in\,[1,4]$) in a
        1.8\Msun star at $X_{\rm c}' = 0.4$ for $(l,m) = (1,1)$. The mode frequencies
        $\omega_{n11}$ were computed with GYRE, adopting the TAR for a uniform
        rotation at 20\percent of the critical rotation velocity in the Roche
        formalism. The dashed lines are placed at the mode frequencies and the
        black dots mark the position of the radial nodes of each mode. The solid
        black line is the Brunt-V\"ais\"al\"a frequency, the dotted line is the
        dipole-mode Lamb frequency $S_{l = 1}$. The effect of rotation is
          only taken into account in the computation of the mode frequencies and
          not in the equilibrium model \citep[][for a thorough justification of such an approach in gravity-mode asteroseismology]{aerts2018}. 
        }}  
    \label{fig:mode_cavity}
\end{figure}

\section{Deviance functions}
\label{AppendixA2}
\comm{Below we show the behaviour of the deviance functions used for determining the confidence intervals from ensemble modelling as described in Section~\ref{sec:err_est}.}
\begin{figure}
    \centering
    \includegraphics[width = \columnwidth]{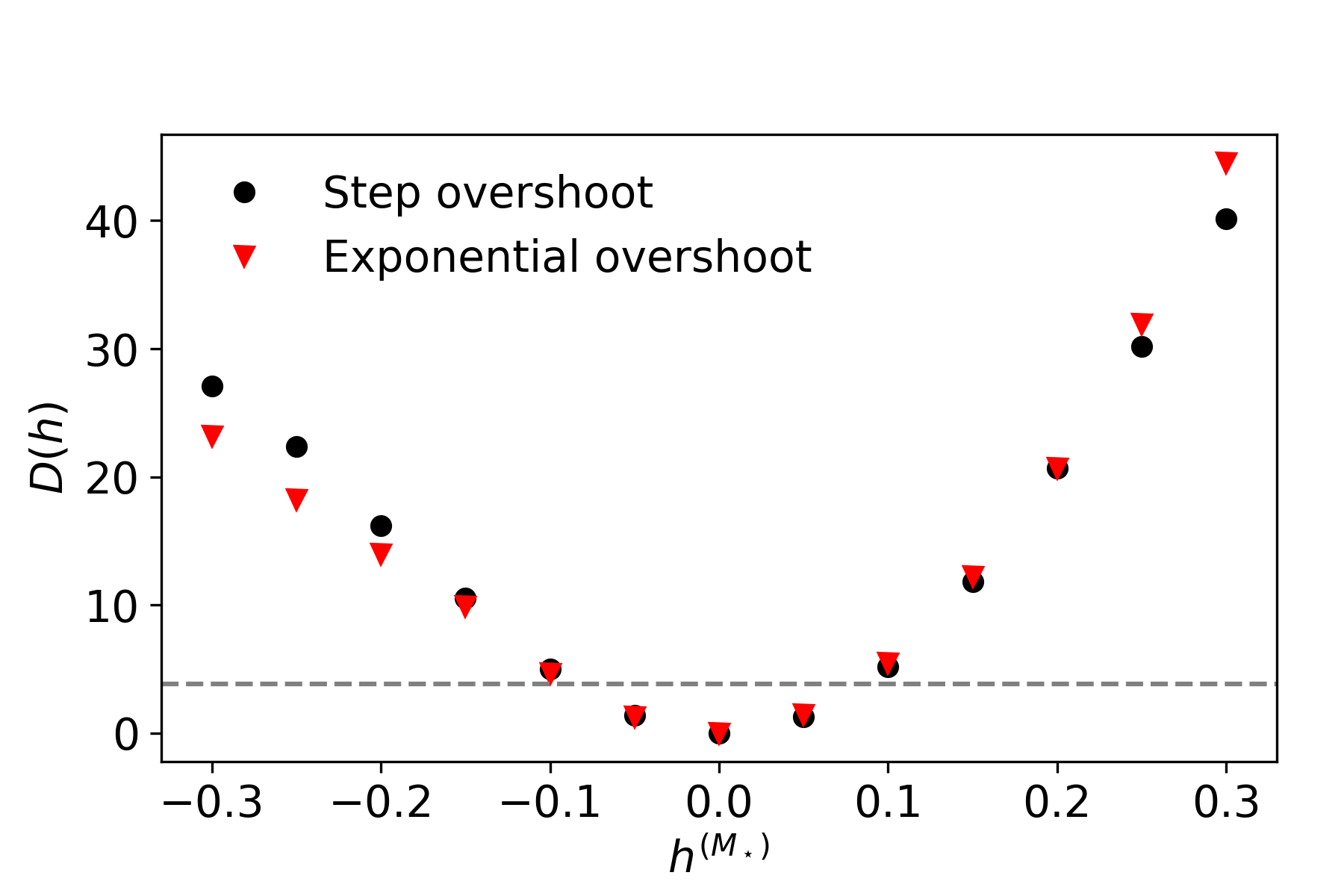}
    \caption{deviance function $D(h)$, where $h$ is a `perturbation' in $M_\star$ of the best model. The dashed grey line at \cas{3.84} indicates the critical value of the $\chi_1^2$ distribution. }
    \label{fig:Dh_M}
\end{figure}

\begin{figure}
    \centering
    \includegraphics[width = \columnwidth]{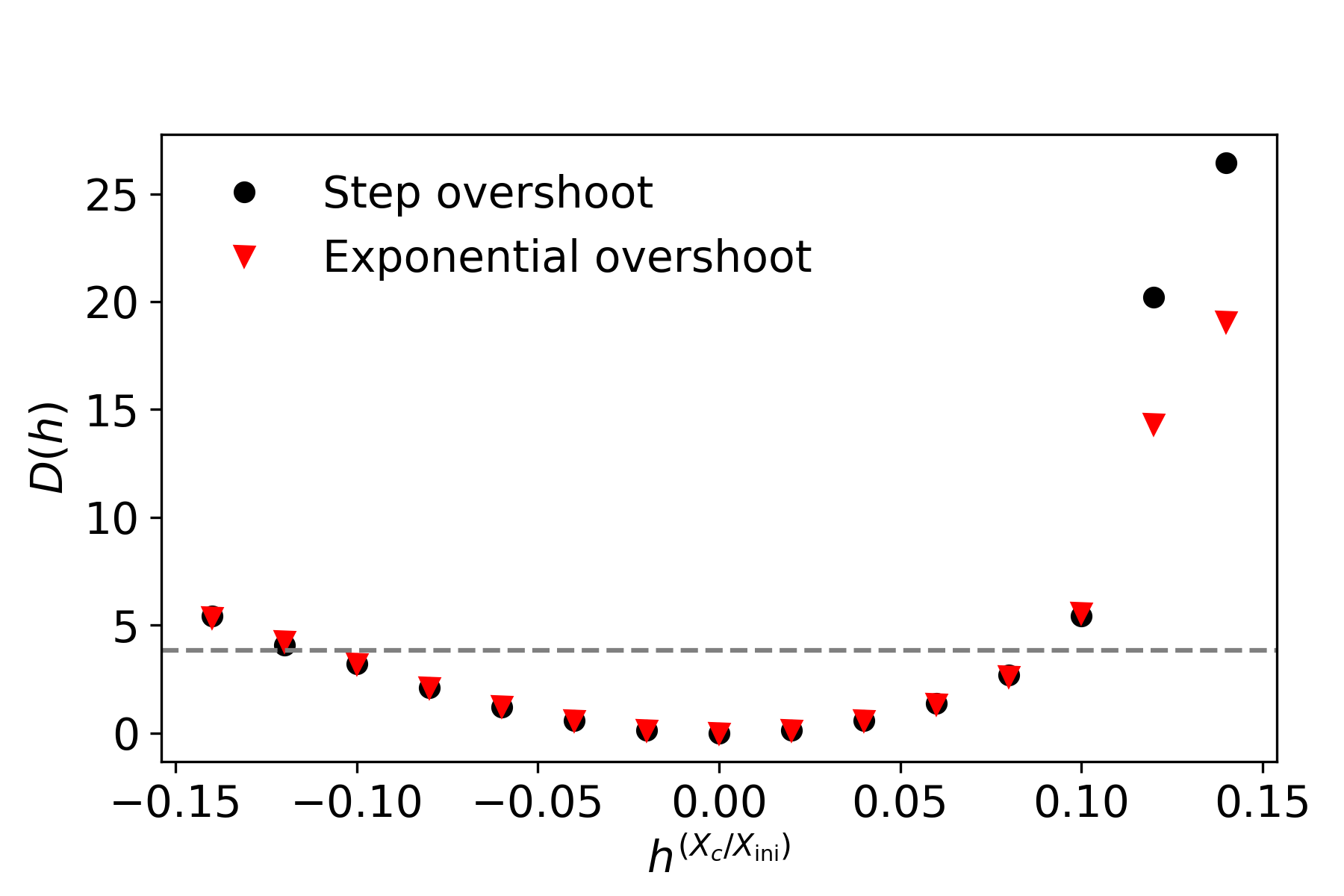}
    \caption{Same as Fig.\,\ref{fig:Dh_M}, but for $X_{\rm c}'$.}
    \label{fig:Dh_xc}
\end{figure}

\begin{figure}
    \centering
    \includegraphics[width = \columnwidth]{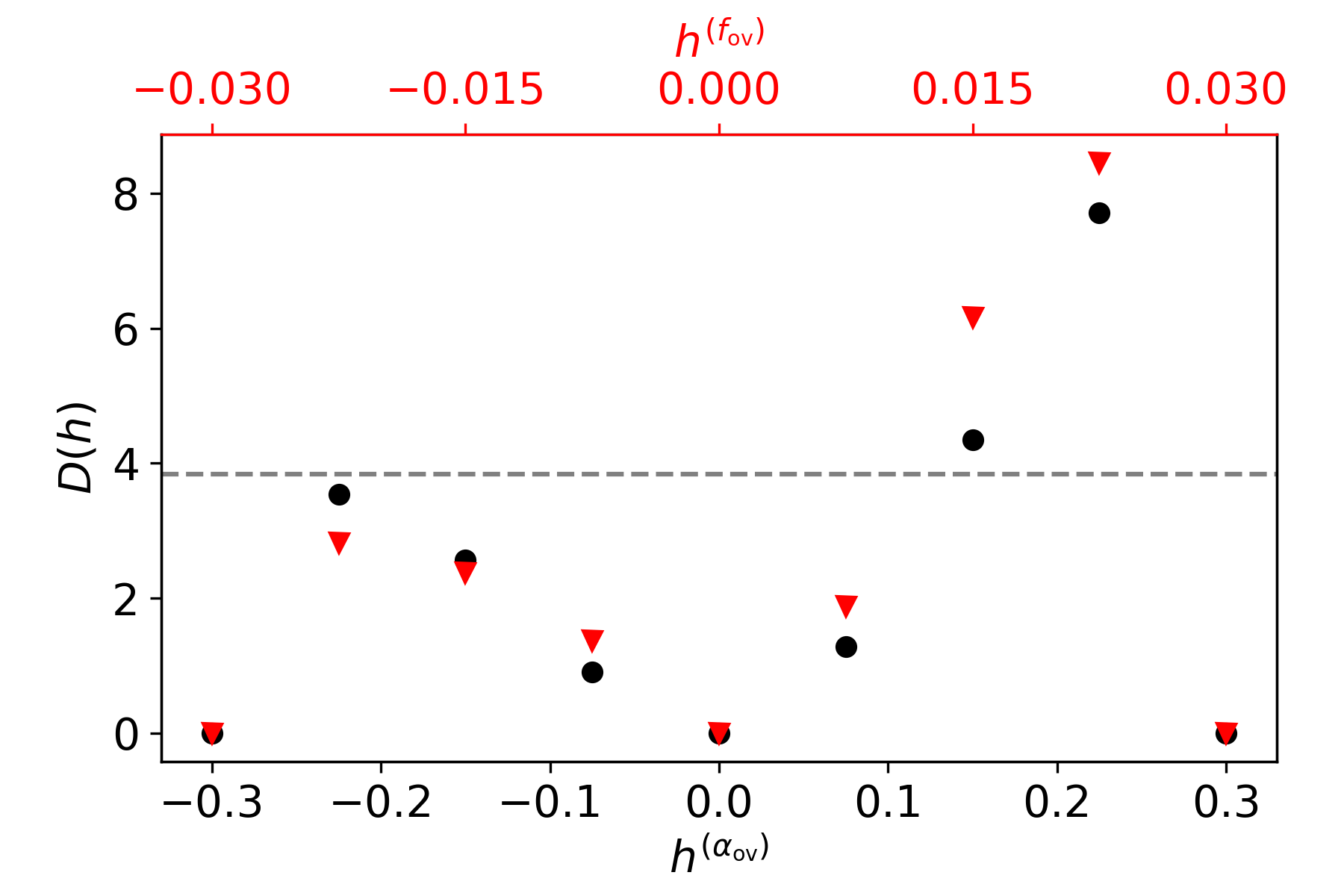}
    \caption{Same as Fig.\,\ref{fig:Dh_M}, but for overshoot. \newline \newline\,}
    \label{fig:Dh_ov}
\end{figure}

\begin{figure}
    \centering
    \includegraphics[width = \columnwidth]{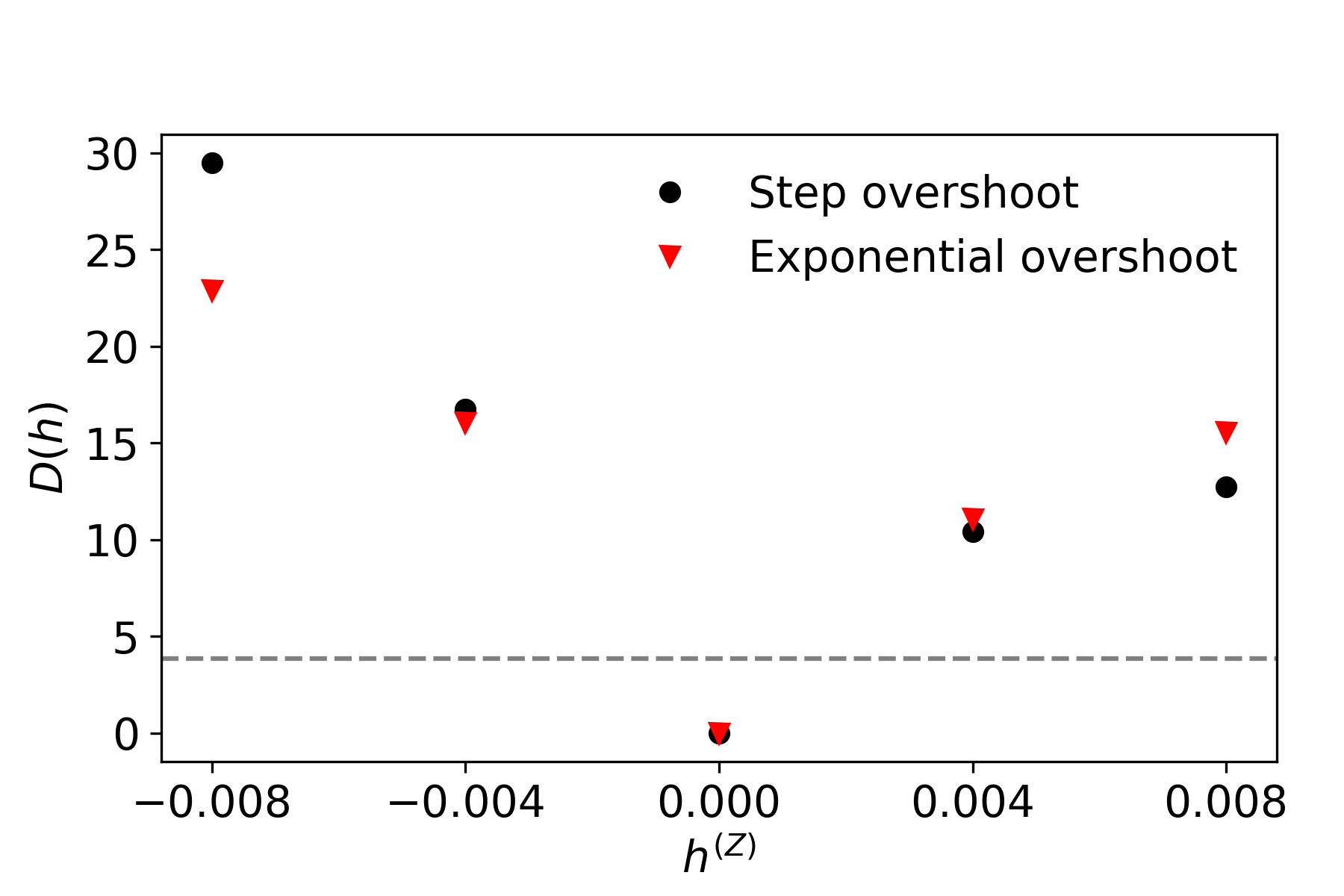}
    \caption{Same as Fig.\,\ref{fig:Dh_M}, but for $Z$.}
    \label{fig:Dh_z}
\end{figure}

\section{Influence of the mixing length parameter} 
\label{appendix:alpha_MLT} 

In this appendix we present the differences in the theoretical prediction of $\Pi_0$,
$T_{\rm eff}$, and $\log g$ caused by the use of 
a different value of $\alpha_{\rm MLT}$ in the computation of stellar models
for a high-mass and low-mass \gDor star. 
\begin{figure}
    \centering
    \includegraphics[width = \columnwidth]{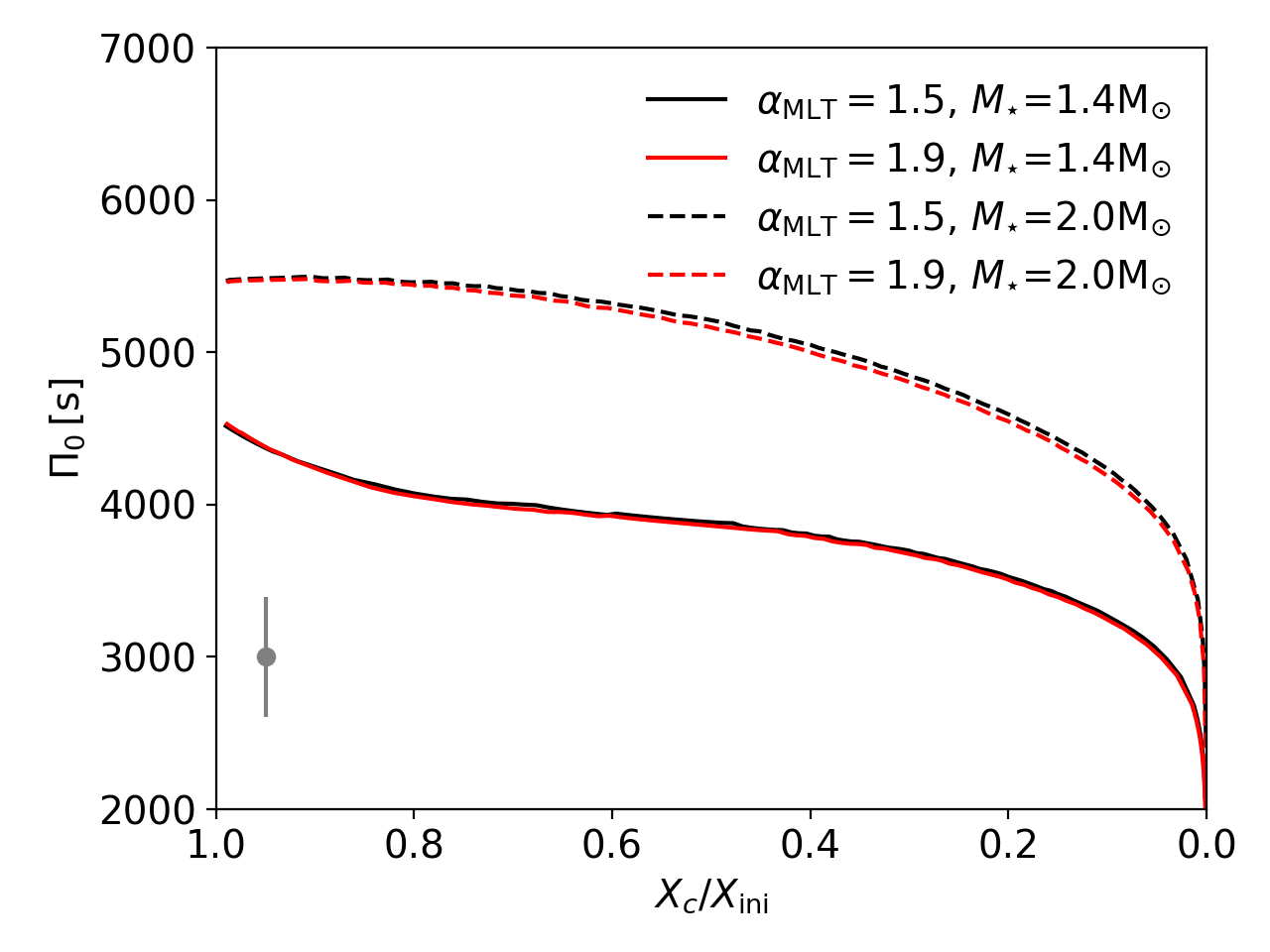}
    \caption{Evolution of $\Pi_0$ for different values of the mixing length
      parameter $\alpha_{\rm MLT}$. The average uncertainty on $\Pi_0$ (in grey)
      for the stars in our sample is also plotted for comparison.}
    \label{fig:Pi0}
\end{figure}
\begin{figure}
    \centering
    \includegraphics[width = \columnwidth]{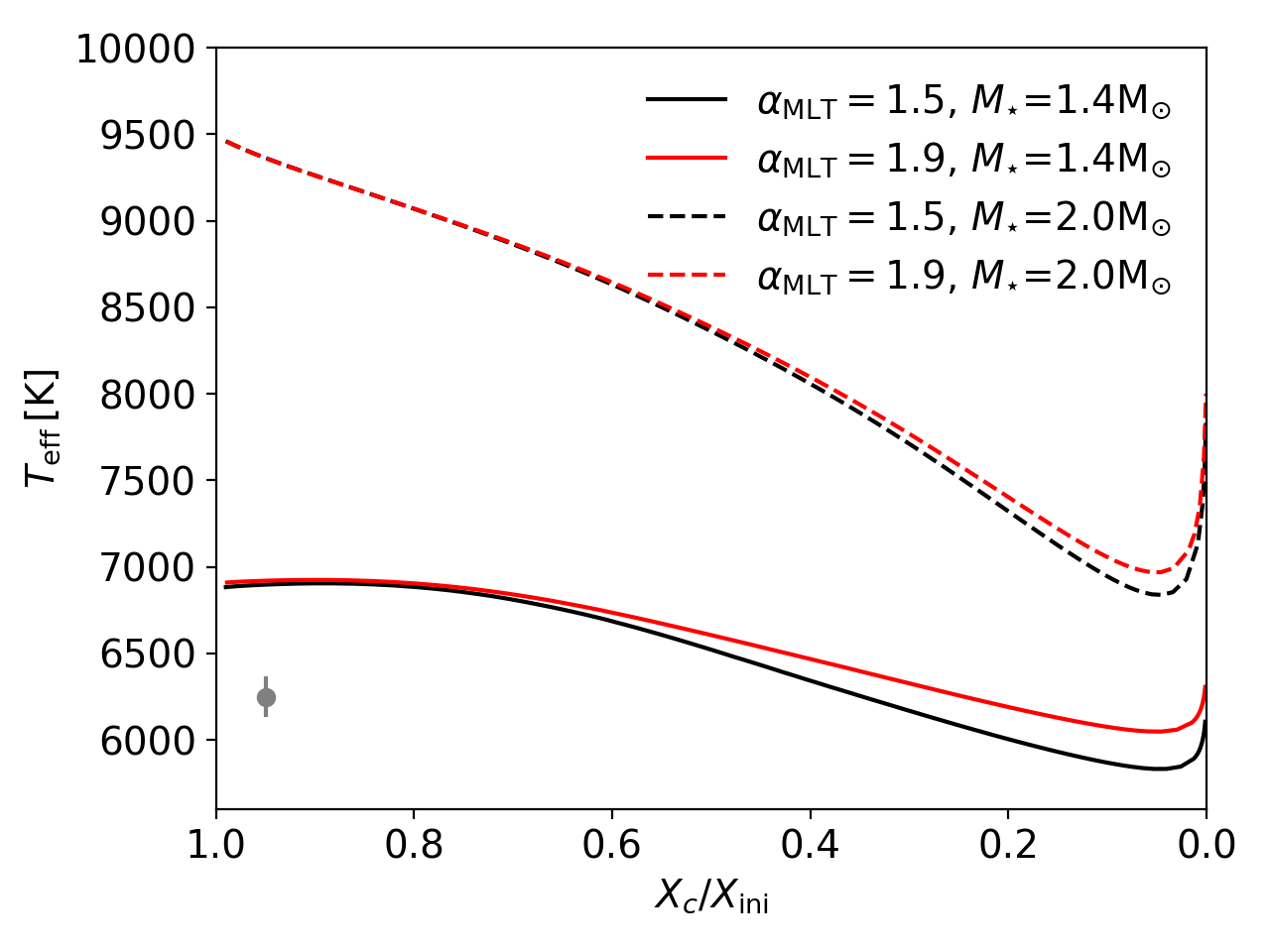}
    \caption{Same as Fig.\,\ref{fig:Pi0}, but for $T_{\rm eff}$.}
    \label{fig:Teff}
\end{figure}
\begin{figure}
    \centering
    \includegraphics[width = \columnwidth]{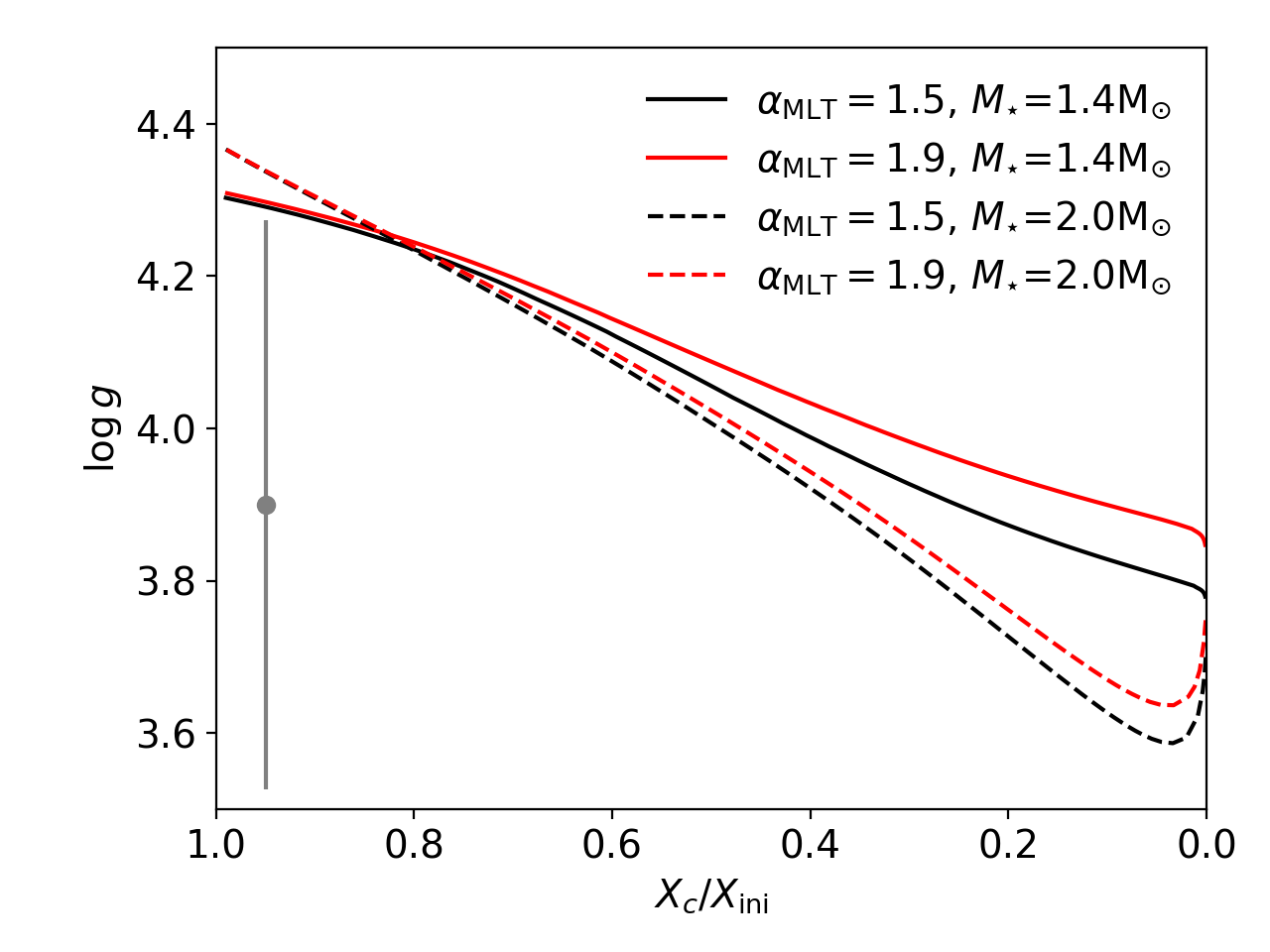}
    \caption{Same as Fig.\,\ref{fig:Pi0}, but for $\log g$.}
    \label{fig:logg}
\end{figure}

\section{Maximum likelihood estimates of the stellar parameters}
\label{AppendixA1}

\comm{Below we present the maximum likelihood estimates 
described in Sections\,\ref{sec:Mahalanobis} and \ref{sec:err_est} 
for each of the five stellar parameters estimated for all the 37 stars in our sample,
along with the observed value of $\Pi_0$.}

\begin{table*}
    \centering
    \begin{tabular}{ccccccccccc}
    \hline
KIC ID & $\Pi_0^{\rm (obs)}$[s] & $M_\star[M_\odot]$ & $X_{\rm c}'$ & $\alpha_{\rm ov}$ & $f_{\rm ov}$  & $Z$  & $ M_{\rm cc}/M_\star$ & $R_{\rm cc}/R_\star$  \\
\hline
\it{KIC2710594 } & 4301 $\substack{+175 \\ -170}$ & 2.00 &  0.14 $\substack{+0.10 \\ -0.12}$ & 0.075 $\substack{+0.150 \\ -0.065}$ & - & 0.018 & 0.082 $\substack{+0.013 \\ -0.019}$  &  0.049 $\substack{+0.011 \\ -0.013}$  \\
          &      & 1.65 &  0.29 $\substack{+0.10 \\ -0.12}$ & - & 0.0300 $\substack{+0.0150 \\ -0.0290}$ & 0.010  & 0.104 $\substack{+0.014 \\ -0.028}$  &  0.053 $\substack{+0.020 \\ -0.012}$ \\
\it{KIC3448365 } & 4237 $\substack{+91 \\ -91}$ & 1.55 &  0.49 $\substack{+0.10 \\ -0.12}$ & 0.150 $\substack{+0.150 \\ -0.140}$ & - & 0.014 & 0.103 $\substack{+0.020 \\ -0.024}$  &  0.073 $\substack{+0.010 \\ -0.012}$  \\
          &      & 1.60 &  0.43 $\substack{+0.10 \\ -0.12}$ & - & 0.0075 $\substack{+0.0150 \\ -0.0065}$ & 0.014  & 0.100 $\substack{+0.021 \\ -0.018}$  &  0.070 $\substack{+0.010 \\ -0.014}$ \\
KIC4846809 & 4144 $\substack{+198 \\ -212}$ & 1.50 &  0.48 $\substack{+0.10 \\ -0.12}$ & 0.225 $\substack{+0.150 \\ -0.215}$ & - & 0.010 & 0.109 $\substack{+0.017 \\ -0.032}$  &  0.074 $\substack{+0.011 \\ -0.012}$ \\
          &      & 1.50 &  0.47 $\substack{+0.10 \\ -0.12}$ & - & 0.0150 $\substack{+0.0150 \\ -0.0140}$ & 0.010  & 0.106 $\substack{+0.023 \\ -0.028}$  &  0.073 $\substack{+0.011 \\ -0.014}$ \\
\it{KIC5114382 } & 4315 $\substack{+129 \\ -122}$ & 1.50 &  0.74 $\substack{+0.10 \\ -0.12}$ & 0.150 $\substack{+0.150 \\ -0.140}$ & - & 0.014 & 0.104 $\substack{+0.025 \\ -0.044}$  &  0.088 $\substack{+0.014 \\ -0.014}$  \\
          &      & 1.50 &  0.72 $\substack{+0.10 \\ -0.12}$ & - & 0.0075 $\substack{+0.0150 \\ -0.0065}$ & 0.014  & 0.099 $\substack{+0.028 \\ -0.035}$  &  0.086 $\substack{+0.014 \\ -0.011}$ \\
KIC5522154 & 4738 $\substack{+57 \\ -42}$ & 1.50 &  0.74 $\substack{+0.10 \\ -0.12}$ & 0.300 $\substack{+0.150 \\ -0.290}$ & - & 0.018 & 0.112 $\substack{+0.011 \\ -0.060}$  &  0.089 $\substack{+0.009 \\ -0.015}$ \\
          &      & 1.50 &  0.75 $\substack{+0.10 \\ -0.12}$ & - & 0.0300 $\substack{+0.0150 \\ -0.0290}$ & 0.018  & 0.116 $\substack{+0.010 \\ -0.062}$  &  0.090 $\substack{+0.010 \\ -0.015}$ \\
KIC5708550 & 4709 $\substack{+339 \\ -311}$ & 1.70 &  0.47 $\substack{+0.10 \\ -0.12}$ & 0.225 $\substack{+0.150 \\ -0.215}$ & - & 0.018 & 0.110 $\substack{+0.015 \\ -0.023}$  &  0.071 $\substack{+0.012 \\ -0.012}$ \\
          &      & 1.70 &  0.50 $\substack{+0.10 \\ -0.12}$ & - & 0.0225 $\substack{+0.0150 \\ -0.0215}$ & 0.018  & 0.115 $\substack{+0.015 \\ -0.026}$  &  0.072 $\substack{+0.013 \\ -0.013}$ \\
KIC5788623 & 3960 $\substack{+679 \\ -622}$ & 1.40 &  0.66 $\substack{+0.10 \\ -0.12}$ & 0.150 $\substack{+0.150 \\ -0.140}$ & - & 0.010 & 0.098 $\substack{+0.028 \\ -0.051}$  &  0.084 $\substack{+0.014 \\ -0.016}$ \\
          &      & 1.40 &  0.64 $\substack{+0.10 \\ -0.12}$ & - & 0.0075 $\substack{+0.0150 \\ -0.0065}$ & 0.010  & 0.092 $\substack{+0.032 \\ -0.041}$  &  0.081 $\substack{+0.014 \\ -0.013}$ \\
KIC6468146 & 4243 $\substack{+156 \\ -156}$ & 1.80 &  0.21 $\substack{+0.10 \\ -0.12}$ & 0.300 $\substack{+0.150 \\ -0.290}$ & - & 0.010 & 0.096 $\substack{+0.014 \\ -0.024}$  &  0.049 $\substack{+0.020 \\ -0.011}$ \\
          &      & 1.85 &  0.18 $\substack{+0.10 \\ -0.12}$ & - & 0.0225 $\substack{+0.0150 \\ -0.0215}$ & 0.010  & 0.094 $\substack{+0.018 \\ -0.024}$  &  0.047 $\substack{+0.019 \\ -0.015}$ \\
\it{KIC6468987 } & 4591 $\substack{+100 \\ -95}$ & 1.75 &  0.44 $\substack{+0.10 \\ -0.12}$ & 0.075 $\substack{+0.150 \\ -0.065}$ & - & 0.018 & 0.103 $\substack{+0.015 \\ -0.016}$  &  0.071 $\substack{+0.010 \\ -0.011}$  \\
          &      & 1.60 &  0.55 $\substack{+0.10 \\ -0.12}$ & - & 0.0300 $\substack{+0.0150 \\ -0.0290}$ & 0.014  & 0.122 $\substack{+0.012 \\ -0.035}$  &  0.076 $\substack{+0.013 \\ -0.011}$ \\
KIC6678174 & 4766 $\substack{+1004 \\ -877}$ & 1.95 &  0.29 $\substack{+0.10 \\ -0.12}$ & 0.225 $\substack{+0.150 \\ -0.215}$ & - & 0.018 & 0.100 $\substack{+0.014 \\ -0.019}$  &  0.056 $\substack{+0.015 \\ -0.013}$ \\
          &      & 2.00 &  0.25 $\substack{+0.10 \\ -0.12}$ & - & 0.0150 $\substack{+0.0150 \\ -0.0140}$ & 0.018  & 0.096 $\substack{+0.017 \\ -0.018}$  &  0.053 $\substack{+0.014 \\ -0.017}$ \\
KIC6935014 & 4497 $\substack{+438 \\ -424}$ & 1.60 &  0.51 $\substack{+0.10 \\ -0.12}$ & 0.300 $\substack{+0.150 \\ -0.290}$ & - & 0.014 & 0.115 $\substack{+0.012 \\ -0.031}$  &  0.074 $\substack{+0.012 \\ -0.011}$ \\
          &      & 1.60 &  0.51 $\substack{+0.10 \\ -0.12}$ & - & 0.0225 $\substack{+0.0150 \\ -0.0215}$ & 0.014  & 0.115 $\substack{+0.016 \\ -0.029}$  &  0.073 $\substack{+0.012 \\ -0.013}$ \\
KIC6953103 & 5035 $\substack{+693 \\ -622}$ & 1.55 &  0.99 $\substack{+0.00 \\ -0.12}$ & 0.300 $\substack{+0.150 \\ -0.290}$ & - & 0.018 & 0.110 $\substack{+0.018 \\ -0.070}$  &  0.104 $\substack{+0.006 \\ -0.030}$ \\
          &      & 1.55 &  0.99 $\substack{+0.00 \\ -0.12}$ & - & 0.0300 $\substack{+0.0150 \\ -0.0290}$ & 0.018  & 0.111 $\substack{+0.019 \\ -0.059}$  &  0.105 $\substack{+0.006 \\ -0.023}$ \\
KIC7023122 & 4780 $\substack{+57 \\ -42}$ & 1.65 &  0.73 $\substack{+0.10 \\ -0.12}$ & 0.150 $\substack{+0.150 \\ -0.140}$ & - & 0.018 & 0.116 $\substack{+0.021 \\ -0.029}$  &  0.091 $\substack{+0.013 \\ -0.012}$ \\
          &      & 1.65 &  0.74 $\substack{+0.10 \\ -0.12}$ & - & 0.0150 $\substack{+0.0150 \\ -0.0140}$ & 0.018  & 0.119 $\substack{+0.020 \\ -0.031}$  &  0.092 $\substack{+0.013 \\ -0.012}$ \\
KIC7365537 & 4723 $\substack{+42 \\ -42}$ & 1.35 &  0.99 $\substack{+0.00 \\ -0.12}$ & 0.225 $\substack{+0.150 \\ -0.215}$ & - & 0.010 & 0.091 $\substack{+0.028 \\ -0.065}$  &  0.101 $\substack{+0.010 \\ -0.034}$ \\
          &      & 1.35 &  0.99 $\substack{+0.00 \\ -0.12}$ & - & 0.0225 $\substack{+0.0150 \\ -0.0215}$ & 0.010  & 0.094 $\substack{+0.028 \\ -0.061}$  &  0.102 $\substack{+0.010 \\ -0.034}$ \\
KIC7380501 & 4045 $\substack{+255 \\ -240}$ & 1.95 &  0.12 $\substack{+0.10 \\ -0.12}$ & 0.010 $\substack{+0.150 \\ -0.000}$ & - & 0.018 & 0.077 $\substack{+0.014 \\ -0.052}$  &  0.049 $\substack{+0.008 \\ -0.023}$ \\
          &      & 1.95 &  0.11 $\substack{+0.10 \\ -0.11}$ & - & 0.0010 $\substack{+0.0150 \\ -0.0000}$ & 0.018  & 0.076 $\substack{+0.015 \\ -0.049}$  &  0.048 $\substack{+0.008 \\ -0.021}$ \\
KIC7434470 & 4271 $\substack{+71 \\ -71}$ & 1.45 &  0.71 $\substack{+0.10 \\ -0.12}$ & 0.150 $\substack{+0.150 \\ -0.140}$ & - & 0.014 & 0.097 $\substack{+0.026 \\ -0.049}$  &  0.085 $\substack{+0.013 \\ -0.014}$ \\
          &      & 1.45 &  0.72 $\substack{+0.10 \\ -0.12}$ & - & 0.0150 $\substack{+0.0150 \\ -0.0140}$ & 0.014  & 0.101 $\substack{+0.026 \\ -0.051}$  &  0.086 $\substack{+0.013 \\ -0.014}$ \\
\it{KIC7583663 } & 4240 $\substack{+150 \\ -146}$ & 1.55 &  0.51 $\substack{+0.10 \\ -0.12}$ & 0.150 $\substack{+0.150 \\ -0.140}$ & - & 0.014 & 0.104 $\substack{+0.021 \\ -0.024}$  &  0.074 $\substack{+0.010 \\ -0.012}$  \\
          &      & 1.55 &  0.49 $\substack{+0.10 \\ -0.12}$ & - & 0.0075 $\substack{+0.0150 \\ -0.0065}$ & 0.014  & 0.100 $\substack{+0.023 \\ -0.019}$  &  0.073 $\substack{+0.010 \\ -0.012}$ \\
KIC7939065 & 4243 $\substack{+57 \\ -57}$ & 1.60 &  0.51 $\substack{+0.10 \\ -0.12}$ & 0.010 $\substack{+0.150 \\ -0.000}$ & - & 0.014 & 0.097 $\substack{+0.022 \\ -0.013}$  &  0.076 $\substack{+0.010 \\ -0.010}$ \\
          &      & 1.45 &  0.58 $\substack{+0.10 \\ -0.12}$ & - & 0.0300 $\substack{+0.0150 \\ -0.0290}$ & 0.010  & 0.121 $\substack{+0.011 \\ -0.053}$  &  0.081 $\substack{+0.012 \\ -0.012}$ \\
KIC8364249 & 4370 $\substack{+156 \\ -156}$ & 1.75 &  0.36 $\substack{+0.10 \\ -0.12}$ & 0.010 $\substack{+0.150 \\ -0.000}$ & - & 0.018 & 0.094 $\substack{+0.017 \\ -0.012}$  &  0.067 $\substack{+0.009 \\ -0.013}$ \\
          &      & 1.60 &  0.46 $\substack{+0.10 \\ -0.12}$ & - & 0.0150 $\substack{+0.0150 \\ -0.0140}$ & 0.014  & 0.107 $\substack{+0.021 \\ -0.023}$  &  0.070 $\substack{+0.011 \\ -0.014}$ \\
\it{KIC8375138 } & 4151 $\substack{+92 \\ -61}$ & 1.45 &  0.47 $\substack{+0.10 \\ -0.12}$ & 0.300 $\substack{+0.150 \\ -0.290}$ & - & 0.010 & 0.110 $\substack{+0.012 \\ -0.040}$  &  0.072 $\substack{+0.012 \\ -0.010}$  \\
          &      & 1.45 &  0.51 $\substack{+0.10 \\ -0.12}$ & - & 0.0300 $\substack{+0.0150 \\ -0.0290}$ & 0.010  & 0.117 $\substack{+0.012 \\ -0.046}$  &  0.074 $\substack{+0.012 \\ -0.011}$ \\

    \hline
    \end{tabular}
    \caption{MLE for each star in our sample, based on the model grids. The
      confidence intervals were derived from treating the sample as an
      ensemble. The uncertainties on $M_\star$ and $Z$ are 0.1\Msun and 0.004,
      respectively. When the confidence interval of $X_{\rm c}'$ or $\alpha_{\rm
        ov}/f_{\rm ov}$ extends beyond physical values, it is truncated at the
      edge of grid as is explained in Section~\ref{sec:err_est}. Uncertainties
      on the mass and size of the convective core are derived from the extreme
      values from the models that lay within the confidence intervals of the
      estimated parameters. \cas{Stars in italic have observed Rossby modes. The
        second column lists the observed values of $\Pi_0$ computed by 
\citet{VanReeth2018}.}}

    \label{tab:MLE_Grid}
\end{table*}

\begin{table*}
    \centering
    \begin{tabular}{ccccccccccc}
    \hline
            KIC ID & $\Pi_0$[s] & $M_\star$[\Msun] & $X_{\rm c}'$ & $\alpha_{\rm ov}$ & $f_{\rm ov}$  & $Z$  & $ M_{\rm cc}/M_\star$ & $R_{\rm cc}/R_\star$  \\
    \hline
    
KIC8645874 & 4525 $\substack{+14 \\ -14}$ & 2.00 &  0.22 $\substack{+0.10 \\ -0.12}$ & 0.075 $\substack{+0.150 \\ -0.065}$ & - & 0.018 & 0.090 $\substack{+0.012 \\ -0.015}$  &  0.055 $\substack{+0.010 \\ -0.013}$ \\
          &      & 1.85 &  0.28 $\substack{+0.10 \\ -0.12}$ & - & 0.0150 $\substack{+0.0150 \\ -0.0140}$ & 0.014  & 0.099 $\substack{+0.020 \\ -0.019}$  &  0.057 $\substack{+0.015 \\ -0.017}$ \\
KIC8836473 & 4101 $\substack{+141 \\ -141}$ & 1.80 &  0.23 $\substack{+0.10 \\ -0.12}$ & 0.010 $\substack{+0.150 \\ -0.000}$ & - & 0.014 & 0.087 $\substack{+0.017 \\ -0.013}$  &  0.058 $\substack{+0.009 \\ -0.014}$ \\
          &      & 1.60 &  0.33 $\substack{+0.10 \\ -0.12}$ & - & 0.0150 $\substack{+0.0150 \\ -0.0140}$ & 0.010  & 0.100 $\substack{+0.021 \\ -0.022}$  &  0.063 $\substack{+0.014 \\ -0.017}$ \\
\it{KIC9480469 } & 4581 $\substack{+112 \\ -119}$ & 1.55 &  0.66 $\substack{+0.10 \\ -0.12}$ & 0.300 $\substack{+0.150 \\ -0.290}$ & - & 0.014 & 0.121 $\substack{+0.011 \\ -0.044}$  &  0.086 $\substack{+0.012 \\ -0.012}$  \\
          &      & 1.65 &  0.65 $\substack{+0.10 \\ -0.12}$ & - & 0.0010 $\substack{+0.0150 \\ -0.0000}$ & 0.018  & 0.103 $\substack{+0.024 \\ -0.012}$  &  0.084 $\substack{+0.012 \\ -0.009}$ \\
KIC9595743 & 4313 $\substack{+156 \\ -156}$ & 1.60 &  0.51 $\substack{+0.10 \\ -0.12}$ & 0.075 $\substack{+0.150 \\ -0.065}$ & - & 0.014 & 0.102 $\substack{+0.017 \\ -0.018}$  &  0.076 $\substack{+0.010 \\ -0.010}$ \\
          &      & 1.60 &  0.52 $\substack{+0.10 \\ -0.12}$ & - & 0.0075 $\substack{+0.0150 \\ -0.0065}$ & 0.014  & 0.105 $\substack{+0.023 \\ -0.019}$  &  0.077 $\substack{+0.010 \\ -0.013}$ \\
KIC9751996 & 4364 $\substack{+7 \\ -7}$ & 1.80 &  0.27 $\substack{+0.10 \\ -0.12}$ & 0.150 $\substack{+0.150 \\ -0.140}$ & - & 0.014 & 0.094 $\substack{+0.018 \\ -0.018}$  &  0.057 $\substack{+0.013 \\ -0.016}$ \\
          &      & 1.95 &  0.20 $\substack{+0.10 \\ -0.12}$ & - & 0.0010 $\substack{+0.0150 \\ -0.0000}$ & 0.018  & 0.086 $\substack{+0.016 \\ -0.014}$  &  0.055 $\substack{+0.008 \\ -0.016}$ \\
KIC10467146 & 4158 $\substack{+849 \\ -764}$ & 1.75 &  0.19 $\substack{+0.10 \\ -0.12}$ & 0.225 $\substack{+0.150 \\ -0.215}$ & - & 0.010 & 0.090 $\substack{+0.016 \\ -0.022}$  &  0.049 $\substack{+0.017 \\ -0.014}$ \\
          &      & 1.95 &  0.13 $\substack{+0.10 \\ -0.12}$ & - & 0.0075 $\substack{+0.0150 \\ -0.0065}$ & 0.014  & 0.081 $\substack{+0.018 \\ -0.026}$  &  0.047 $\substack{+0.012 \\ -0.017}$ \\
KIC11080103 & 4752 $\substack{+1245 \\ -1032}$ & 1.60 &  0.88 $\substack{+0.10 \\ -0.12}$ & 0.150 $\substack{+0.150 \\ -0.140}$ & - & 0.018 & 0.109 $\substack{+0.024 \\ -0.051}$  &  0.098 $\substack{+0.015 \\ -0.015}$ \\
          &      & 1.60 &  0.88 $\substack{+0.10 \\ -0.12}$ & - & 0.0150 $\substack{+0.0150 \\ -0.0140}$ & 0.018  & 0.112 $\substack{+0.023 \\ -0.043}$  &  0.098 $\substack{+0.015 \\ -0.015}$ \\
KIC11099031 & 5035 $\substack{+141 \\ -156}$ & 1.60 &  0.51 $\substack{+0.10 \\ -0.12}$ & 0.300 $\substack{+0.150 \\ -0.290}$ & - & 0.018 & 0.112 $\substack{+0.011 \\ -0.030}$  &  0.073 $\substack{+0.011 \\ -0.010}$ \\
          &      & 1.60 &  0.53 $\substack{+0.10 \\ -0.12}$ & - & 0.0300 $\substack{+0.0150 \\ -0.0290}$ & 0.018  & 0.116 $\substack{+0.011 \\ -0.033}$  &  0.073 $\substack{+0.012 \\ -0.011}$ \\
KIC11294808 & 3917 $\substack{+495 \\ -453}$ & 1.85 &  0.12 $\substack{+0.10 \\ -0.12}$ & 0.075 $\substack{+0.150 \\ -0.065}$ & - & 0.014 & 0.078 $\substack{+0.014 \\ -0.055}$  &  0.047 $\substack{+0.011 \\ -0.021}$ \\
          &      & 1.70 &  0.17 $\substack{+0.10 \\ -0.12}$ & - & 0.0150 $\substack{+0.0150 \\ -0.0140}$ & 0.010  & 0.086 $\substack{+0.020 \\ -0.021}$  &  0.048 $\substack{+0.015 \\ -0.017}$ \\
KIC11456474 & 3974 $\substack{+354 \\ -339}$ & 1.50 &  0.38 $\substack{+0.10 \\ -0.12}$ & 0.225 $\substack{+0.150 \\ -0.215}$ & - & 0.010 & 0.101 $\substack{+0.016 \\ -0.028}$  &  0.065 $\substack{+0.012 \\ -0.012}$ \\
          &      & 1.50 &  0.37 $\substack{+0.10 \\ -0.12}$ & - & 0.0150 $\substack{+0.0150 \\ -0.0140}$ & 0.010  & 0.099 $\substack{+0.022 \\ -0.025}$  &  0.065 $\substack{+0.012 \\ -0.015}$ \\
KIC11721304 & 4356 $\substack{+608 \\ -566}$ & 1.55 &  0.59 $\substack{+0.10 \\ -0.12}$ & 0.150 $\substack{+0.150 \\ -0.140}$ & - & 0.014 & 0.108 $\substack{+0.022 \\ -0.028}$  &  0.081 $\substack{+0.011 \\ -0.010}$ \\
          &      & 1.55 &  0.61 $\substack{+0.10 \\ -0.12}$ & - & 0.0150 $\substack{+0.0150 \\ -0.0140}$ & 0.014  & 0.112 $\substack{+0.022 \\ -0.031}$  &  0.082 $\substack{+0.011 \\ -0.012}$ \\
KIC11754232 & 4426 $\substack{+28 \\ -28}$ & 1.70 &  0.43 $\substack{+0.10 \\ -0.12}$ & 0.075 $\substack{+0.150 \\ -0.065}$ & - & 0.014 & 0.102 $\substack{+0.016 \\ -0.016}$  &  0.071 $\substack{+0.011 \\ -0.012}$ \\
          &      & 1.70 &  0.44 $\substack{+0.10 \\ -0.12}$ & - & 0.0075 $\substack{+0.0150 \\ -0.0065}$ & 0.014  & 0.105 $\substack{+0.021 \\ -0.017}$  &  0.072 $\substack{+0.011 \\ -0.015}$ \\
KIC11826272 & 4172 $\substack{+410 \\ -396}$ & 1.60 &  0.30 $\substack{+0.10 \\ -0.12}$ & 0.300 $\substack{+0.150 \\ -0.290}$ & - & 0.010 & 0.101 $\substack{+0.013 \\ -0.026}$  &  0.057 $\substack{+0.017 \\ -0.011}$ \\
          &      & 1.75 &  0.25 $\substack{+0.10 \\ -0.12}$ & - & 0.0075 $\substack{+0.0150 \\ -0.0065}$ & 0.014  & 0.090 $\substack{+0.019 \\ -0.016}$  &  0.056 $\substack{+0.012 \\ -0.016}$ \\
\it{KIC11907454 } & 4203 $\substack{+59 \\ -57}$ & 1.45 &  0.69 $\substack{+0.10 \\ -0.12}$ & 0.075 $\substack{+0.150 \\ -0.065}$ & - & 0.014 & 0.088 $\substack{+0.023 \\ -0.039}$  &  0.082 $\substack{+0.012 \\ -0.011}$  \\
          &      & 1.45 &  0.70 $\substack{+0.10 \\ -0.12}$ & - & 0.0075 $\substack{+0.0150 \\ -0.0065}$ & 0.014  & 0.091 $\substack{+0.030 \\ -0.040}$  &  0.083 $\substack{+0.014 \\ -0.011}$ \\
KIC11917550 & 4101 $\substack{+311 \\ -297}$ & 1.55 &  0.48 $\substack{+0.10 \\ -0.12}$ & 0.010 $\substack{+0.150 \\ -0.000}$ & - & 0.014 & 0.092 $\substack{+0.023 \\ -0.013}$  &  0.073 $\substack{+0.009 \\ -0.009}$ \\
          &      & 1.55 &  0.49 $\substack{+0.10 \\ -0.12}$ & - & 0.0010 $\substack{+0.0150 \\ -0.0000}$ & 0.014  & 0.093 $\substack{+0.024 \\ -0.013}$  &  0.073 $\substack{+0.009 \\ -0.010}$ \\
KIC11920505 & 4214 $\substack{+368 \\ -354}$ & 1.55 &  0.55 $\substack{+0.10 \\ -0.12}$ & 0.010 $\substack{+0.150 \\ -0.000}$ & - & 0.014 & 0.094 $\substack{+0.024 \\ -0.013}$  &  0.077 $\substack{+0.011 \\ -0.008}$ \\
          &      & 1.55 &  0.56 $\substack{+0.10 \\ -0.12}$ & - & 0.0010 $\substack{+0.0150 \\ -0.0000}$ & 0.014  & 0.096 $\substack{+0.026 \\ -0.013}$  &  0.078 $\substack{+0.011 \\ -0.009}$ \\
\it{KIC12066947 } & 4185 $\substack{+58 \\ -58}$ & 1.40 &  0.94 $\substack{+0.05 \\ -0.12}$ & 0.010 $\substack{+0.150 \\ -0.000}$ & - & 0.010 & 0.065 $\substack{+0.046 \\ -0.030}$  &  0.087 $\substack{+0.020 \\ -0.015}$  \\
          &      & 1.40 &  0.94 $\substack{+0.05 \\ -0.12}$ & - & 0.0010 $\substack{+0.0150 \\ -0.0000}$ & 0.010  & 0.068 $\substack{+0.047 \\ -0.025}$  &  0.088 $\substack{+0.021 \\ -0.015}$ \\
   \hline
    \end{tabular}
    \contcaption{}
\end{table*}

\bsp	
\label{lastpage}
 
\end{document}